\documentclass[a4paper,11pt]{article}

\usepackage{latexsym}
\usepackage{amsfonts} 
\usepackage[mathscr]{euscript}
\usepackage{amsmath,mathrsfs,bm,amssymb,color, amsthm}
\usepackage{authblk}

\usepackage{bbm}
\usepackage{enumitem}
\usepackage{stmaryrd}
\usepackage{nicefrac}
\usepackage{xcolor}
\usepackage{bm}

\usepackage{ulem}
\usepackage{comment}

\author[1]{Tadahiro Miyao}

\affil[1]{Department of Mathematics, Hokkaido University

Sapporo 060-0810 Japan

E-mail: miyao@math.sci.hokudai.ac.jp}

\title{\bf On renormalized Hamiltonian nets
}
\date{}

\newcommand{\one}{{\mathchoice {\rm 1\mskip-4mu l} {\rm 1\mskip-4mu l}
{\rm 1\mskip-4.5mu l} {\rm 1\mskip-5mu l}}}
\newcommand{\h}{\mathfrak{H}}

\newcommand{\D}{\mathrm{dom}}

\newcommand{\Fock}{\mathfrak{F}}

\newcommand{\dG}{d\Gamma}

\newcommand{\la}{\langle}
\newcommand{\ra}{\rangle}

\newcommand{\slim}{\mbox{$\mathrm{s}$-$\displaystyle\lim_{n\to\infty}$}}

\newcommand{\BbbR}{\mathbb{R}}
\newcommand{\BbbN}{\mathbb{N}}

\newcommand{\BbbC}{\mathbb{C}}

\newcommand{\vepsilon}{\varepsilon}
\newcommand{\vphi}{\varphi}
\newcommand{\Pt}{P_{\mathrm{tot}}}
\newcommand{\Pf}{P_{\mathrm{f}}}

\newcommand{\Hf}{H_{\mathrm{f}}}

\newcommand{\Cone}{\mathfrak{P}}

\newcommand{\no}{\nonumber \\}

\newcommand{\Bb}{\mathbb{B}^d_{\mathrm{b}}}

\newcommand{\bs}{\boldsymbol}

\newcommand{\MI}{\mathfrak{M}^{\iota}}
\newcommand{\MU}{\mathfrak{M}^{\upsilon}}
\newcommand{\HI}{\mathfrak{H}^{\iota}}
\newcommand{\HU}{\mathfrak{H}^{\upsilon}}
\newcommand{\PI}{\mathfrak{P}^{\iota}}
\newcommand{\PU}{\mathfrak{P}^{\upsilon}}
\newcommand{\OI}{\Omega^{\iota}}
\newcommand{\OU}{\Omega^{\upsilon}}

\usepackage[margin=1in]{geometry}

\begin{document}

\newtheorem{define}{Definition}[section]
\newtheorem{Thm}[define]{Theorem}
\newtheorem{Prop}[define]{Proposition}
\newtheorem{lemm}[define]{Lemma}
\newtheorem{rem}[define]{Remark}
\newtheorem{assum}{Condition}
\newtheorem{example}{Example}
\newtheorem{coro}[define]{Corollary}

\theoremstyle{definition}
\newtheorem*{Proof}{Proof}

\maketitle

\begin{abstract}
We propose  an abstract framework describing      energy-renormalized Hamiltonians in terms of  local algebras.
  Within the framework, we examine the positivity improvingness of the semigroup
generated by the renormalized Hamiltonian.  As examples, we discuss the  renormalized Nelson Hamiltonian and the renormalized Nelson Hamiltonian at  fixed total momentum.
 The characteristic features of our approach are as follows:
 (i) in contrast with the probabilistic approach in the Schr\"odinger representation,  
 our method works well  in the Fock representation;  and  (ii)  the  method
 covers the massless case. 
\begin{flushleft}
{\bf Mathematics Subject Classification (2010).} 
\end{flushleft}
Primary:  47A63, 47D08,  46L60;
Secondary: 47N50, 81T10
\begin{flushleft}
{\bf
Keywords. 
} 
\end{flushleft}
 Local algebras; Tomita-Takesaki theory;
Ergodic semigroups; Positivity improving semigroups; Nelson model;  Energy renormalization.
\end{abstract}

\section{Introduction}\label{Introduction}
Let us consider the Nelson model
which describes a system of a  single quantum mechanical particle coupled with a scalar bose field:
\begin{align}
 H_{\mathrm{Nelson},\kappa}=-\frac{1}{2}\Delta_x-V-g\int_{\BbbR^3}dk\frac{1_{B_{\kappa}}(k)}{\sqrt{\vepsilon(k)}}
\big(e^{ik\cdot x}a(k)+e^{-ik\cdot x}a(k)^*\big)+\Hf-E_{\kappa}, \label{NelsonHamiV}
\end{align}
where $\Delta_x$ is the Laplacian on $L^2(\BbbR^3, dx)$,  $V$ is a confining potential and $x$ is the position operator of the particle.
In the remainder of this section, we assume that $V\in L^2(\BbbR^3, dx)+L^{\infty}(\BbbR^3, dx)$.
The operator $ H_{\mathrm{Nelson}, \kappa}$ acts on $L^2(\BbbR^3,  dx)\otimes \Fock(L^2(\BbbR^3))$, where 
$\Fock(\mathfrak{h})$ is the Fock space over $\mathfrak{h}$:
$
\Fock(\mathfrak{h})=\bigoplus_{n=0}^{\infty} \mathfrak{h}^{\otimes_{\mathrm{s}} n}
$. Here,  $\mathfrak{h}^{\otimes_{\mathrm{s}}n}$ is the $n$
-fold symmetric tensor product with $\mathfrak{h}^{\otimes_{\mathrm{s}}0}=\BbbC$.
$a(k)$ and $a(k)^*$ are annihilation- and creation operators which satisfy the standard commutation relations:
\begin{align}
[a(k), a(k')^*] =\delta(k-k'),\ \ \ [a(k), a(k')]=0.\label{CCRs}
\end{align}
The field momentum operator $\Pf=(P_{\mathrm{f}, 1}, P_{\mathrm{f}, 2}, P_{\mathrm{f}, 3})$ is defined by 
\begin{align}
P_{\mathrm{f}, i}=\int_{\BbbR^3} dk k_i a(k)^* a(k),\ \ \ i=1,2,3. \label{MomOp}
\end{align}
The field energy $\Hf$ is
\begin{align}
\Hf=\int_{\BbbR^3} dk\varepsilon(k)a(k)^*a(k),\ \ \ \varepsilon(k)=\sqrt{k^2+m^2}. \label{FieldOp}
 \end{align} 
 The energy renormalization $E_{\kappa}$ is defined by
\begin{align}
E_{\kappa}=-g^2\int_{\BbbR^3} dk\frac{1_{B_{\kappa}}(k)}{\varepsilon(k) \{\varepsilon(k)+k^2/2\}}, \label{ReEn}
\end{align}
where $B_{\kappa}$ is the  ball of radius $\kappa$ centered at the origin and $1_{B_{\kappa}}$ is the indicator function of the set $B_{\kappa}$. Notice   that $E_{\kappa} \to -\infty$ as $\kappa\to \infty$.
$g$ is the  coupling strength between the particle and the field.  Without loss of generality, we may assume that $g>0$.
The mass of bosons is denoted by $m\ge 0$. 
By applying the Kato-Rellich theorem  \cite[Theorem X.12]{ReSi2}, we can prove that $H_{{\rm Nelson},  \kappa}$ is self-adjoint and bounded from below.
We emphasize here that our results  in the present paper
can cover the massless case: $m=0$.
\medskip

Let $\Pt$ be  the total momentum operator: $\Pt=-i\nabla_x+\Pf$. Here, the symbol $\nabla_x$ is the
standard nabla: $\nabla_x=(\frac{\partial}{\partial x_1}, \frac{\partial}{\partial x_2}, \frac{\partial}{\partial x_3})$.\footnote{
To be precise, $P_{\mathrm{tot}, j}$ is essentially self-adjoint for each $j=1, 2, 3$. In what follows,  we denote its closure by the same symbol.
}  
Assume  that $V\equiv 0$.
Then the total momentum is conserved, i.e., $H_{\mathrm{Nelson}, \kappa}^{V\equiv 0}$, the Nelson Hamiltonian (\ref{NelsonHamiV}) with $V\equiv 0$,  commutes with $\Pt$. Therefore, one has the decomposition
\begin{align}
\mathscr{U} H_{\mathrm{Nelson}, \kappa}^{V\equiv 0}\mathscr{U}^{-1}=\int^{\oplus}_{\BbbR^3}H_{\kappa}(P)dP,
\end{align}
where the unitary operator $\mathscr{U}$ is given by $\mathscr{U}=\mathscr{F}\exp(ix\cdot \Pt)$. Here, $\mathscr{F}$ denotes the Fourier transformation. For each $P\in \BbbR^3$, 
$H_{\kappa}(P)$ is defined by 
\begin{align}
H_{\kappa}(P)=\frac{1}{2}(P-\Pf)^2-g\int_{\BbbR^3} dk \frac{1_{B_{\kappa}}(k)}{\sqrt{\vepsilon(k)}}
(a(k)+a(k)^*) +\Hf-E_{\kappa}.
\end{align} 
The operator $H_{\kappa}(P)$ is called the Hamiltonian at a  fixed total momentum $P$. Remark that $H_{\kappa}(P) $
is a self-adjoint operator acting on $\Fock(L^2(\BbbR^3))$, bounded from below.
\medskip

We wish to remove the   ultraviolet cutoff, namely, we are interested in the model with   $\kappa= \infty$.
At a first glance, such a  limiting Hamiltonian cannot be defined mathematically because  
the form factor $1/\sqrt{\varepsilon(k)}$  is not square-integrable.\footnote{
Recall that the following operators
$$
\int_{\BbbR^3} dk f(k)^*a(k),\ \ \ \int_{\BbbR^3} dk f(k)a(k)^*
$$
are mathematically meaningful only if $f$ is square-integrable.
}
Surprisingly, Nelson proves the following result:
\begin{Thm}\label{NelsonEx}
\begin{itemize}
\item[\rm (i)] There exists a self-adjoint operator $H_{\mathrm{Nelson}}$,  bounded from below,  such that 
$H_{\mathrm{Nelson}, \kappa}$ converges to $H_{\mathrm{Nelson}}$ in the strong resolvent sense as $\kappa\to \infty$.
\item[\rm (ii)] There exists a self-adjoint operator $H_{\mathrm{ren}}(P)$,   bounded from below,  such that 
$H_{\kappa}(P)$ converges to $H_{\mathrm{ren}}(P)$ in the strong resolvent sense as $\kappa\to \infty$.
Furthermore, it holds that 
\begin{align}
\mathscr{U}H_{\mathrm{Nelson}}^{V\equiv 0}\mathscr{U}^{-1}=\int^{\oplus}_{\BbbR^3}H_{\mathrm{ren}}(P)dP.
\end{align}
\end{itemize}
\end{Thm}
\begin{Proof}
See \cite{Nelson} for (i) and see \cite[Proposition 4.7]{Miyao5} for (ii). Note that, in these papers, the condition $m>0$ is assumed, however, we can straightforwardly extend the proofs to the case where $m=0$. \qed 
\end{Proof}

An important point of Theorem \ref{NelsonEx} is that the renormalized Hamiltonians can be defined without changing the representation space;  in compensation for this, we need  the infinite energy renormalization: $E_{\kappa}\approx -\infty$.
Because the Nelson model possesses such a unique property, the model  has been actively studied,   see, e.g., \cite{AH,Ammari,A. Arai,Dam, DM, GW,GHL,HHS,LS, LoMS,Pizzo,Pizzo2,Sasaki,Spohn};  mathematical analysis of the model has been known to be hard since the model is indirectly defined through the infinite energy renormalization.
\medskip

In order to explain our purpose of the present paper, let us focus  on the Hamiltonian at  fixed total momentum.
Let $\mathbb{B}^3$ be the Borel sets of $\BbbR^3$.
For each $\Lambda \in \mathbb{B}^3$, bounded, we define the  Hamiltonian with an ultraviolet cutoff $\Lambda$ by 
\begin{align}
H^{\iota}(\Lambda)=\frac{1}{2}(P-P_{\mathrm{f}, \Lambda})^2-g\int_{\BbbR^3}dk\frac{1_{\Lambda}(k)}{\sqrt{\varepsilon(k)}}
(a(k)+a(k)^*)+H_{\mathrm{f}, \Lambda}-E(\Lambda),
\end{align}
where
\begin{align}
P_{\mathrm{f}, \Lambda}&=\int_{\Lambda} dk k a(k)^*a(k), \label{DPL}\\
H_{\mathrm{f}, \Lambda}&=\int_{\Lambda} dk \varepsilon(k)a(k)^* a(k), \label{DHL}\\
E(\Lambda)&= -g^2\int_{\Lambda} \frac{dk}{\varepsilon(k) \{\varepsilon(k)+k^2/2\}}. \label{DEng}
\end{align}
Because $\Lambda$ is bounded, we can choose $\kappa$ so that $\Lambda \subset B_{\kappa}$.
With this choice, we  introduce 
\begin{align}
H_{\kappa}^{\upsilon}(\Lambda^c)=\frac{1}{2}P_{\mathrm{f}, \Lambda^c}^2-g\int_{\BbbR^3}dk\frac{1_{B_{\kappa} \setminus \Lambda}(k)}{\sqrt{\varepsilon(k)}}
(a(k)+a(k)^*)+H_{\mathrm{f}, \Lambda^c}-E(B_{\kappa} \setminus\Lambda),
\end{align}
where $\Lambda^c$ stands for the complement of $\Lambda$.
Then we have  the following algebraic relation:
\begin{align}
H_{\kappa}(P)=H^{\iota}(\Lambda)+W(\Lambda)+H_{\kappa}^{\upsilon}(\Lambda^c), \label{KappaDec2}
\end{align}
where 
\begin{align}
W(\Lambda)=-(P-P_{\mathrm{f}, \Lambda})\cdot P_{\mathrm{f}, \Lambda^c}. \label{Gomi}
\end{align}
Taking the limit $\kappa\to \infty$, we formally obtain
\begin{align}
H_{\mathrm{ren}}(P)=H^{\iota}(\Lambda)+W(\Lambda)+H^{\upsilon}(\Lambda^c). \label{KappaDec3}
\end{align}
 Note that the mathematical justification of (\ref{KappaDec3}) is examined in Section \ref{NelsonNet}. 
Relations of this kind often appear in quantum statistical mechanics, see, e.g.,  \cite{BR2, Simon}; 
from  those algebraic relations,
many useful results can be derived; for instance,
  it is possible to characterize the KMS states in terms of operator algebras. Therefore, it is natural to explore the renormalized Nelson model by using the ideas in quantum statistical mechanics.
A main purpose in this study is to construct an abstract framework describing energy-renormalized  models 
and to study its properties from a  viewpoint of  quantum statistical mechanics; especially, we examine the semigroups generated by the renormalized Hamiltonians in terms of operator algebras.
\medskip

We will mainly  study whether the semigroups $e^{-\beta H_{\mathrm{ren}}(P)}$ and $e^{-\beta H_{\mathrm{Nelson}}}$ improve the positivity. 
The positivity improvingness of $e^{-\beta H_{\mathrm{ren}}(P)}$ in the Fock representation was first conjectured by Fr\"ohlich \cite{JFroehlich1,JFroehlich2}. In \cite{MM}, Matte-M\o ller has succeeded to 
prove the positivity improvingness of $e^{-\beta H_{\mathrm{Nelson}}}$, not $e^{-\beta H(P)}$, in the Schr\"odinger representation. Then the author solved the Fr\"ohlich conjecture in \cite{Miyao5}.
\medskip

There are two approaches to the problem. One is Matte-M\o ller's method \cite{MM} based on the path integral formula; their method
is applicable in the Schr\"odinger representation.
The other approach is established by the author in \cite{Miyao3,Miyao4,Miyao5}; his  method originates from Fr\"ohlich's pioneering works
\cite{JFroehlich1,JFroehlich2} and is effective in the Fock representation. As discussed  in Section \ref{NelsonNetII}, these two methods complement each other and have specific advantages.
Our algebraic approach in this  paper provides a general framework of the author\rq{}s works.
As we will perform, the present method naturally covers the massless case. 
Moreover, we will prove that the positivity improvingness of the semigroup  $e^{-\beta H_{\mathrm{Nelson}}}$, not $e^{-\beta H_{\rm ren}(P)}$,  in the {\it Fock representation}.  As far as we know, this result is new and 
provide information on the renormalized Nelson Hamiltonian which is   different from the one obtained  in the Schr\"odinger representation.
\medskip

The organization of the present paper is as follows:
In Section \ref{DefRe}, we introduce the concept of   renormalized Hamiltonian nets  which is a generalization of (\ref{KappaDec3}). Then we  illustrate an abstract theory of the positivity improving  semigroups associated with  the renormalized Hamiltonian net. In Section \ref{Proofs}, we prove the results in Section \ref{DefRe}.
Section \ref{NelsonNet} is devoted to give an example of the renormalized Hamiltonian net associated with the Nelson Hamiltonian at  fixed total momentum: $H_{\mathrm{ren}}(P)$. In Section \ref{NelsonNetII}, we examine the renormalized Hamiltonian net associated with the Nelson Hamiltonian with   confining  potential:    $H_{\mathrm{Nelson}}$. 
In Appendices \ref{UsefulT} and \ref{DefTens}, we give a list of fundamental facts that are used in the main sections. In Appendices \ref{ProofNelsonII} and  \ref{Ext2V}, we prove the positivity improvingness of the semigroup generated by the Hamiltonian $H_{\mathrm{Nelson}, \kappa}$ in the Fock representation. 
This fact is a basic input in Section \ref{NelsonNetII}.

\subsection*{Acknowledgements}
I  thank the Mathematisches Forschungsinstitut Oberwolfach
for its hospitality. I am grateful to J. S.  M\o ller for useful discussions,  to W. Dybalski for helpful comments, and to J. Lampart  for  sending  me  a   draft of  his paper \cite{Lampart}. 
The hospitality support by the Aarhus university is gratefully acknowledged.
This work was partially supported by   KAKENHI 18K0331508,  JSPS bilateral joint research project between Danish and Japan,  and the Research Institute for Mathematical
Sciences, an International Joint Usage/Research Center located in Kyoto
University. 
I am   grateful to   the anonymous referees for the  constructive comments and suggestions,
which helped considerably to improve the presentation of the manuscript.

\section{An abstract theory} \label{DefRe}
In this section, we will construct  an abstract theory of positivity improving semigroups  associated with  renormalized Hamiltonian. In Sections \ref{NelsonNet} and \ref{NelsonNetII}, we will examine the renormalized Nelson Hamiltonians as  important examples.
\subsection{Local structures}\label{QLocal}
\setcounter{equation}{0}

Let $\mathbb{B}^d$ be the Borel sets of $\BbbR^d$ and let $\Bb=\{B\in \mathbb{B}^d\, |\, \mbox{$B$ is bounded and $|B|\neq 0$}\}$,  where $|B|$ is the Lebesgue measure of $B$.

\begin{define}\label{DefMnet}
\upshape Let $\mathfrak{H}$ be a complex  separable   Hilbert space   and let $\mathscr{B}(\h)$ be the set of all bounded operators on $\h$.
 Let $\mathfrak{M}\subset \mathscr{B}(\h)$ be a von Neumann algebra on  $\mathfrak{H}$.
    We say that   $\mathfrak{M}$ admits   a {\it  local structure}  if 
     there exists  a  net $\{\mathfrak{M}_{\Lambda}\}_{\Lambda\in \mathbb{B}^d}$ of 
    von Neumann subalgebras satisfying the following:
\begin{itemize}
\item[(i)] $\mathfrak{M}_{\Lambda}\neq \varnothing$ if and only if $|\Lambda|\neq 0$.
\item[(ii)] If $\Lambda\subset \Lambda'$ and $|\Lambda'\setminus \Lambda| \neq 0$, then $\mathfrak{M}_{\Lambda} \subset \mathfrak{M}_{\Lambda'}$.
\item[(iii)] If $|\Lambda\cap \Lambda'|=0$, then $[\mathfrak{M}_{\Lambda}, \mathfrak{M}_{\Lambda'}]=\{0\}$.
\item[(iv)]  $\mathfrak{M}=\overline{\bigcup_{\Lambda\in \Bb} \mathfrak{M}_{\Lambda}}^{\mathrm{s}}$, where
$\overline{S}^{\mathrm{s}}$ is the closure of $S$ in the strong operator topology. 
\end{itemize}
\end{define}

\begin{define} \upshape
Let $\mathfrak{M}$ be a von Neumann  algebra on $\mathfrak{H}$ admitting a local structure $\{\mathfrak{M}_{\Lambda}\}_{\Lambda\in \mathbb{B}^d}$.
We say that   a net $\{\mathfrak{H}_{\Lambda}\}_{\Lambda\in \mathbb{B}^d}$ of  closed subspaces of $\h$
and a net $\{\Omega_{\Lambda}\}_{\Lambda\in \mathbb{B}^d}$ of unit vectors  are  {\it adapted} to $\{\mathfrak{M}_{\Lambda}\}_{\Lambda\in \mathbb{B}^d}$ if 
   the following conditions are satisfied:
 \begin{itemize}
 \item[(i)] $\mathfrak{M}_{\Lambda}\subset \mathscr{B}(\h_{\Lambda})$. Here, we understand that $\mathfrak{H}_{\BbbR^d}=\mathfrak{H}$ and 
$\mathfrak{M}_{\BbbR^d}=\mathfrak{M}$.

\item[(ii)] $\Omega_{\Lambda}\in \h_{\Lambda}$ is cyclic and separating for $\mathfrak{M}_{\Lambda}$. 
\end{itemize}
\end{define}

In this study, we further  impose the following tensor product structures:
\begin{define}\label{QLS} \upshape
 Let $\mathfrak{M}$  be a von Neumann  algebra on $\mathfrak{H}$ admitting a local structure $\{\mathfrak{M}_{\Lambda}\}_{\Lambda\in \mathbb{B}^d}$.
  Suppose that $\{\mathfrak{H}_{\Lambda}\}_{\Lambda\in \mathbb{B}^d}$ 
and  $\{\Omega_{\Lambda}\}_{\Lambda\in \mathbb{B}^d}$   are   adapted to $\{\mathfrak{M}_{\Lambda}\}_{\Lambda\in \mathbb{B}^d}$.
We say that  $\mathfrak{M}$ is  {\it factorizable} if the following are satisfied:
\begin{itemize}
 \item[(i)] 
 If $\Lambda\subset \Lambda', |\Lambda|\neq 0$ and  $|\Lambda'\setminus  \Lambda|\neq 0$, then $\h_{\Lambda'}=\h_{\Lambda}\otimes \h_{\Lambda'\setminus  \Lambda}$. In particular, 
 $\mathfrak{H}=\mathfrak{H}_{\Lambda}\otimes \mathfrak{H}_{\Lambda^c}$ for  each $\Lambda\in \mathbb{B}^d$ with $ |\Lambda|\neq 0$ and $|\Lambda^c|\neq 0$, where 
$\Lambda^c$ is the complement of $\Lambda$. 
\item[(ii)] Corresponding to the condition (i), $\Omega_{\Lambda}$ can be factorized  as $\Omega_{\Lambda'}=\Omega_{\Lambda}\otimes \Omega_{\Lambda'\setminus  \Lambda}$, 
provided  that $\Lambda\subset \Lambda',  |\Lambda|\neq 0$ and $|\Lambda'\setminus  \Lambda| \neq 0$.
In particular, 
$\Omega=\Omega_{\Lambda}\otimes \Omega_{\Lambda^c}$ for all $\Lambda\in \mathbb{B}^d$ with $ |\Lambda|\neq 0$ and $|\Lambda^c|\neq 0$, where we set $\Omega=\Omega_{\BbbR^d}$.
\item[(iii)] 
 If $\Lambda\subset \Lambda',  |\Lambda|\neq 0$ and  $|\Lambda'\setminus  \Lambda|\neq 0$, then $\mathfrak{M}_{\Lambda'}=\mathfrak{M}_{\Lambda}\otimes \mathfrak{M}_{\Lambda'\setminus  \Lambda}$, where   $\mathfrak{M}_{\Lambda}\otimes \mathfrak{M}_{\Lambda'\setminus   \Lambda}$ indicates the von Neumann tensor product  of $\mathfrak{M}_{\Lambda}$ and $\mathfrak{M}_{\Lambda'\setminus  \Lambda}$.
 In particular, 
$\mathfrak{M}=\mathfrak{M}_{\Lambda}\otimes \mathfrak{M}_{\Lambda^c}$ for all $\Lambda\in \mathbb{B}^d$ with $ |\Lambda|\neq 0$ and $|\Lambda^c|\neq 0$.
\end{itemize}

\end{define}

\begin{define}\label{QLS2} \upshape
 Let $\MI$ be a von Neumann  algebra on  a  Hilbert space   $\HI$  admitting a  local structure $\{\MI_{\Lambda}\}_{\Lambda\in \mathbb{B}^d}$ and   let  $\MU$ be  a  von Neumann  algebra on a   Hilbert space $\HU$ admitting a  local structure    $\{\MU_{\Lambda}\}_{\Lambda\in \mathbb{B}^d}$.
 Suppose that there are  a net $\{\HI_{\Lambda}\}_{\Lambda\in \mathbb{B}^d}$ (resp. $\{\HU_{\Lambda}\}_{\Lambda\in \mathbb{B}^d}$) of closed subspaces of $\HI$ (resp. $\HU$)
and a net $\{\Omega^{\iota}_{\Lambda}\}_{\Lambda\in \mathbb{B}^d}$ (resp. $\{\Omega^{\upsilon}_{\Lambda}\}_{\Lambda\in \mathbb{B}^d}$) of unit vectors  which  are  adapted to $\{\MI_{\Lambda}\}_{\Lambda\in \mathbb{B}^d}$ (resp. $\{\MU_{\Lambda}\}_{\Lambda\in \mathbb{B}^d}$).
We say that  the pair  $(\MI,  \MU)$ is a {\it generalized local system} if the following are satisfied:
\begin{itemize}
\item[(i)] $\MU$ is factorizable.
 \item[(ii)] 
 If $\Lambda\subset \Lambda'$, $\Lambda\in \mathbb{B}^d_{\rm b}$ and  $|\Lambda'\setminus  \Lambda|\neq 0$, then $\HI_{\Lambda'}=\HI_{\Lambda}\otimes \HU_{\Lambda'\setminus  \Lambda}$. 
 In particular, 
 $\HI=\HI_{\Lambda}\otimes \HU_{\Lambda^c}$ for  each $\Lambda\in \mathbb{B}_{\rm b}^d$.
\item[(iii)]  $\Omega^{\iota}_{\Lambda}$ can be factorized  as $\Omega^{\iota}_{\Lambda'}=\Omega^{\iota}_{\Lambda}\otimes \Omega^{\upsilon}_{\Lambda'\setminus  \Lambda}$, 
provided  that $\Lambda\subset \Lambda'$, $\Lambda\in \mathbb{B}^d_{\rm b}$  and $|\Lambda'\setminus  \Lambda| \neq 0$.
In particular, 
$\Omega^{\iota}=\Omega^{\iota}_{\Lambda}\otimes \Omega^{\upsilon}_{\Lambda^c}$ holds, where  we set $\OI=\OI_{\BbbR^d}$.
\item[(iv)] 
 If $\Lambda\subset \Lambda'$, $\Lambda\in \mathbb{B}^d_{\rm b}$  and  $|\Lambda'\setminus  \Lambda|\neq 0$, then $\MI_{\Lambda'}=\MI_{\Lambda}\otimes \MU_{\Lambda'\setminus  \Lambda}$.
 In particular, 
$\MI=\MI_{\Lambda}\otimes \MU_{\Lambda^c}$ for all $\Lambda\in \mathbb{B}_{\rm b}^d$.
\end{itemize}

\end{define}

\begin{rem}{\rm 
Let $\Lambda, \Lambda'\in \mathbb{B}^d$. Suppose that  $\Lambda\subset \Lambda'$, $\Lambda\in \mathbb{B}^d_{\rm b}$ and  $|\Lambda'\setminus  \Lambda| \neq 0$. 
\begin{itemize}
\item[1. ]
We occasionally identify $\MI_{\Lambda}$ with $\MI_{\Lambda} \otimes \one_{\Lambda'\setminus  \Lambda}$, where $\one_{\Lambda\rq{}\setminus \Lambda}$ indicates the identity operator on $\HU_{\Lambda\rq{}\setminus \Lambda}$. In particular, $\MI_{\Lambda}=\MI_{\Lambda} \otimes \one_{\Lambda^c}$. Similarly, we identify $\MU_{\Lambda\rq{}\setminus \Lambda}$ with $\one_{\Lambda} \otimes \MU_{\Lambda\rq{}\setminus \Lambda}$.
\item[2. ]
Let $A$ and $B$ be linear operators on $\HI_{\Lambda}$ and $\HU_{\Lambda'\setminus  \Lambda}$, respectively. We often leave out tensor factors and write 
$A\cong A\otimes \one_{\Lambda'\setminus  \Lambda}$ and $B\cong \one_{\Lambda} \otimes B$.
\end{itemize}
}
\end{rem}

The following proposition is readily confirmed:
\begin{Prop}\label{Simple}
Let $\mathfrak{M}$  be a von Neumann algebra admitting a local structure $\{\mathfrak{M}_{\Lambda}\}_{\Lambda\in \mathbb{B}^d}$. If $\mathfrak{M}$ is factorizable, then
$(\mathfrak{M}, \mathfrak{M})$ is a generalized local system.
\end{Prop}
Proposition \ref{Simple} will be useful in Section \ref{NelsonI}; in Section \ref{NelsonI}, the Nelson Hamiltonian at fixed total momentum is examined  from a view point of a generalized local system; in this case, the local structure  is determined by   the abelian  von Neumann algebras of the second quantization of the multiplication operators in the momentum space.

\subsection{Renormalized Hamiltonian net}

Let $A$ and $B$ be self-adjoint operators, bounded from below.
The form domain of the operator $A$ is denoted by $Q(A)$.
If $Q(A) \cap Q(B)$ is dense, then we can define the form sum of $A$ and $B$, which is denoted by 
$A\dot{+}B$, see, e.g., \cite[Section VIII.6, Example 4]{ReSi1}.
\begin{define}\label{ReNetDef}
{\rm 
Let  $(\MI,  \MU)$ be  a generalized local system.
Let $H$ be a self-adjoint operator acting in $\HI$, bounded from below.
The  {\it   renormalized  Hamiltonian net  associated with $H$} is 
a net of triplets of self-adjoint operators $\{(H^{\iota}(\Lambda), H^{\upsilon}(\Lambda^c), W(\Lambda))\, |\, \Lambda\in \mathbb{B}_{\rm b}^d \}$
such that the following properties are valid:
\begin{itemize}
\item[(i)] $H^{\#}(\Lambda)$ acts on  $\mathfrak{H}^{\#}_{\Lambda}\ (\#=\iota, \upsilon)$, and  $H^{\iota}(\Lambda)$ and $H^{\upsilon}(\Lambda^c)$ are bounded from below
 for all $\Lambda\in \mathbb{B}_{\rm b}^d$.
 
 \item[(ii)] For each $\Lambda\in \Bb$,  $Q(H)=Q(H^{\iota}(\Lambda)) \cap Q(H^{\upsilon}(\Lambda^c))$.
 \item[(iii)]  For each  $\Lambda \in \Bb$,  there exists a self-adjoint operator $W(\Lambda)$ such that $Q(H)\subseteq Q(W(\Lambda))$ and 
 \begin{align}
 H=H^{\iota}(\Lambda)\dot{+} W(\Lambda)\dot{+}H^{\upsilon}(\Lambda^c).
 \end{align}

\end{itemize}
The operator $H$ is called the  {\it renormalized Hamiltonian}. 
For each $\Lambda\in \Bb$, $H^{\iota}(\Lambda)$ is called the {\it  Hamiltonian with an ultraviolet cutoff $\Lambda$.}
}
\end{define}

\subsection{Theorems}
\subsubsection{Preliminaries}
Let $\mathfrak{M}$ be a von Neumann algebra on a complex  separable Hilbert space  $\mathfrak{H}$, and $\Omega$ be a cyclic and separating vector for $\mathfrak{M}$.
We use  $\Delta$ and $J$ to denote  the modular operator and the modular conjugation associated with the pair $\{\mathfrak{M}, \Omega\}$ \cite[Definition 2.5.10]{BR1}.
The Tomita-Takesaki theorem \cite[Theorem 2.5.14]{BR1} tells us that $J\mathfrak{M}J=\mathfrak{M}'$ and $\Delta^{it } \mathfrak{M} \Delta^{-it}=\mathfrak{M} $ for all $t\in \BbbR$, where $\mathfrak{M}'$ is the commutant of $\mathfrak{M}$.
\begin{define}\label{DefPCone}
{\rm 
 Let 
$
\mathcal{P}_0(\mathfrak{M})=\{A JAJ\, |\, A\in \mathfrak{M}\} .
$
The {\it natural cone},  $\Cone$,  associated with the pair $\{\mathfrak{M}, \Omega\}$ is defined by 
$
\Cone=\overline{\mathcal{P}_0(\mathfrak{M}) \Omega},
$
where the bar denotes  the  closure in the norm in $\h$.

}
\end{define}

It is well-known that  $\Cone$ is a self-dual cone in $\h$, that is,
$\Cone=\Cone^{\dagger}$, where $\Cone^{\dagger}$ is the dual cone of $\Cone:
\, \Cone^{\dagger}=\{\vphi\in \h\, |\, \la \vphi|\psi\ra\ge 0\, \forall \psi\in \Cone\}
$.  
If $J\vphi=\vphi$, then $\vphi$ has a unique decomposition $\vphi=\vphi_+-\vphi_-$, where
$\vphi_{\pm} \in \Cone$ and $\vphi_+\perp \vphi_-$.
See, e.g., \cite[Theorem 2.5.28]{BR1} for detail.

The following order structures are important in this study.
\begin{define}
{\rm 
\begin{itemize}
\item[(i)] A vector $\vphi$ is said to be  {\it positive w.r.t. $\Cone$} if $\vphi\in
 \Cone$.  We write this as $\vphi \ge 0$  w.r.t. $\Cone$.

 \item[(ii)] A vector $\vphi \in \Cone$ is called {\it strictly positive
w.r.t. $\Cone$},  whenever $\la \vphi| \psi\ra>0$ for all $\psi\in
\Cone \setminus \{0\}$. We write this as $\vphi>0 $
w.r.t. $\Cone$.

 \item[(iii)]  Let $\h_{\rm real}=\{\vphi\in \h\, |\, J\vphi=\vphi\}$.
 Let $\vphi, \psi\in \h_{\rm real}$. If $\vphi-\psi\in \Cone$, then we write this as $\vphi\ge \psi$ w.r.t. $\Cone$.

\end{itemize} 
}
\end{define}
In  subsequent  sections, we  use the following order preserving  operator inequalities.
\begin{define}{\rm 
Let $A, B\in \mathscr{B}(\h)$.
\begin{itemize}
\item[(i)]  If $A \Cone\subseteq \Cone,$\footnote{
For each subset $\mathfrak{C}\subseteq \h$, $A\mathfrak{C}$ is
	     defined by $A\mathfrak{C}=\{A\vphi\, |\, \vphi\in \mathfrak{C}\}$.
} we then 
write  this as  $A \unrhd 0$ w.r.t. $\Cone$.\footnote{This
 symbol was introduced by Miura \cite{Miura}.} In
	     this case, we say that {\it $A$ preserves the
positivity w.r.t. $\Cone$.}  
\item[(ii)] Suppose that $A\h_{\rm real}\subseteq
 \h_{\rm real}$ and $B\h_{\rm real } \subseteq
	     \h_{\rm real}$. If $(A-B) \Cone\subseteq
	     \Cone$, then we write this as $A \unrhd B$ w.r.t. $\Cone$. 
\item[(iii)] 
We write  $A\rhd 0$ w.r.t. $\Cone$, if  $A\vphi >0$ w.r.t. $\Cone$ for all $\vphi\in
\Cone \setminus \{0\}$. 
 In this case, we say that {\it $A$ improves the
positivity w.r.t. $\Cone$.} 
\end{itemize}
}
\end{define}

\begin{define}
{\rm
Let $A$ be a positive self-adjoint operator such that $e^{-\beta A} \unrhd 0$ w.r.t. $\Cone$ for all $\beta \ge 0$. We say that the semigroup $e^{-\beta A}$ is {\it ergodic  w.r.t.} $\Cone$, if the following condition is satisfied:  For each $\vphi, \psi\in \Cone\setminus \{0\}$, there exists a $\beta \ge 0$ such that $\la \vphi|e^{-\beta A} \psi\ra >0$. Note that $\beta$ could depend on $\vphi$ and $\psi$.
}
\end{define}

As we will prove in Appendix \ref{UsefulT}, the ergodicity is equivalent to the positivity improvingness,  when $\Cone$ is a lattice.

\subsection{Main results}\label{SecMainResult}
Let  $(\MI,  \MU)$ be  a generalized local system. Let $\{(H^{\iota}(\Lambda), H^{\upsilon}(\Lambda^c), W(\Lambda))\}_{\Lambda\in \mathbb{B}_{\rm b}^d}$
 be a  renormalized Hamiltonian net associated with $H$.

Given $\Lambda\in \mathbb{B}^d$ with $\Lambda\neq \BbbR^d$, we denote by   $\Cone^{\iota}_{\Lambda}$ (resp. $\Cone^{\upsilon}_{\Lambda}$)  the natural  cone associated with the pair $\{\MI_{\Lambda}, \Omega^{\iota}_{\Lambda}\}$ (resp. $\{\MU_{\Lambda}, \Omega^{\upsilon}_{\Lambda}\}$). 
 The natural  cone associated with $\{\MI, \OI\}$ (resp. $\{\MU, \OU\}$) is denoted by $\Cone^{\iota}$ (resp. $\Cone^{\upsilon}$), where we set  $\OI=\OI_{\BbbR^d}$ and $\OU=\OU_{\BbbR^d}$.

As we will see in Proposition \ref{BasicPro}, these self-dual cones are related as follows:  Let $\Lambda, \Lambda\rq{}\in \mathbb{B}^d$. If $\Lambda\subset \Lambda'$, $\Lambda\in \mathbb{B}^d_{\rm b}$ and  $|\Lambda'\setminus  \Lambda|\neq 0$, then, corresponding to (ii) of Definition \ref{QLS2}, we have
\begin{align}
\Cone^{\iota}_{\Lambda'}=\Cone^{\iota}_{\Lambda}\otimes \Cone^{\upsilon}_{\Lambda'\setminus  \Lambda},
\end{align}
where the right hand side is defined in Appendix \ref{DefTens}. In particular, 
\begin{align}
\Cone^{\iota}=\Cone^{\iota}_{\Lambda}\otimes \Cone^{\upsilon}_{\Lambda^c}
\end{align}
holds. These properties will play crucial roles in the present paper. 
Remark that  we can define positivities, positivity preservingness, positivity improvingness and ergodicity with respect to $\Cone^{\iota}_{\Lambda}$,  etc. for each $\Lambda\in \mathbb{B}^d$.

In what follows, we assume the following conditions:
\begin{description}
\item[{\bf (A. 1)}] $(W(\Lambda)+i)^{-1}\in \mathfrak{Z}(\MI)$, where $\mathfrak{Z}(\MI)$ is the center  of $\MI:\ \mathfrak{Z}(\MI)=\MI \cap (\MI)\rq{}$.

\item[{\bf (A. 2)} ]$(\Delta^{\iota})
^{it} W(\Lambda) \subseteq W(\Lambda) (\Delta^{\iota})^{it}$ for all $t\in \BbbR$, where $\Delta^{\iota} $
indicates the modular operator associated with the pair $\{\MI, \OI\}$.

\item[{\bf (A. 3)}] For each $\Lambda \in \mathbb{B}_{\rm b}^d$, $e^{-\beta H^{\iota}(\Lambda)} \unrhd 0$ w.r.t. $\PI_{\Lambda}$ and $e^{-\beta H^{\upsilon}(\Lambda^c)} \unrhd 0$ w.r.t. $\PU_{\Lambda^c}$ for all $\beta \ge 0$.
\end{description}

The following theorem characterizes the ergodicity of the semigroup generated by the renormalized Hamiltonian $H$.

\begin{Thm}\label{MainTh1}
Assume {\bf (A. 1)}, {\bf (A. 2)} and {\bf (A. 3)}.
Given $\Lambda\in \mathbb{B}^d_{\rm b}$, let $L(\Lambda)=H^{\iota}(\Lambda)+H^{\upsilon}(\Lambda^c)$.
The following conditions are equivalent:
\begin{itemize}
\item[{\rm (i)}] The semigroup $e^{-\beta H}$ is ergodic  w.r.t. $\PI$.
\item[{\rm (ii)}] For each $\vphi, \psi\in \PI\setminus \{0\}$, there exist $\beta \ge 0$ and $\Lambda\in \Bb$
such that 
$\la \vphi| e^{-\beta L(\Lambda)}\psi\ra  >0$. 

\end{itemize}
\end{Thm}

We remark that there are some interesting similarities between Theorem \ref{MainTh1} and the characterization of the KMS state by the  Gibbs condition in the quantum statistical mechanics, 
see, e.g.,  \cite{Araki}, \cite[Theorem 6.2.18]{ BR2}.

For concrete applications to the nonrelativistic quantum field theory,  there is a more convenient theorem.
To  state it, we have to introduce the following condition. 
\begin{description}
\item[{\bf (A. 4)}] There exists  a net  $\{\omega^{\upsilon}_{\Lambda}\}_{\Lambda\in \mathbb{B}^d}$ of normalized vectors
satisfying the following:
\begin{itemize}
\item[(i)] For every $\Lambda\in \mathbb{B}^d$, $\omega^{\upsilon}_{\Lambda} \ge 0$ w.r.t. $\PU_{\Lambda}$ and there is a constant $\gamma>0$ independent of $\Lambda$ such that  $\la \omega_{\Lambda}^{\upsilon}|\Omega_{\Lambda}^{\upsilon}\ra\ge \gamma$.
\item[(ii)] If $\Lambda \subset \Lambda', |\Lambda|\neq 0$ and  $|\Lambda'\setminus  \Lambda|\neq 0$, then
$\omega^{\upsilon}_{\Lambda'}=\omega^{\upsilon}_{\Lambda}\otimes \omega^{\upsilon}_{\Lambda'\setminus  \Lambda}$. Recall (ii) of Definition of \ref{QLS}, here.
\item[(iii)] For each $\Lambda\in \Bb$, it holds that $\one^{\iota}_{\Lambda} \otimes |\omega^{\upsilon}_{\Lambda^c}\ra\la \omega^{\upsilon}_{\Lambda^c}| \unlhd \one^{\iota}$ w.r.t. $\PI$, where
 $\one_{\Lambda}^{\iota}$ (resp. $\one^{\iota}$) is the identity operator on $\HI_{\Lambda}$ (resp. $\HI$). 
\end{itemize}
In general, $\omega^{\upsilon}_{\Lambda}$ is different from $\Omega_{\Lambda}^{\upsilon}$.
\end{description}

\begin{rem}\upshape
The main  role of $\OU_{\Lambda}$ is to generate the self-dual cone $\PI_{\Lambda}$. In contrast, the  main role of $\omega^{\upsilon}_{\Lambda}$ is to translate (ii) of Theorem \ref{MainTh1} to (ii) of Theorem \ref{MainTh2Ori} which is more convenient for applications.
\end{rem}

\begin{Thm}\label{MainTh2Ori}
Assume {\bf (A. 1)}, {\bf (A. 2)},   {\bf (A. 3)}  and {\bf (A. 4)}.
The following conditions are equivalent:
\begin{itemize}
\item[{\rm (i)}] $e^{-\beta H}$ is ergodic  w.r.t. $\PI$.
\item[{\rm (ii)}] $e^{-\beta H^{\iota}(\Lambda)} $  is ergodic w.r.t. $\PI_{\Lambda}$ for all $\Lambda\in \Bb$.
\end{itemize}
\end{Thm}

To translate the ergodicity to the positivity improvingness, we need the following condition.

\begin{description}
\item[{\bf (A. 5)}] For each $\Lambda \in \mathbb{B}^d$ and $\vphi, \psi\in \Cone^{\iota}_{\Lambda}$,  it holds that $\vphi\wedge \psi\ge 0$ w.r.t. $\Cone^{\iota}_{\Lambda}$, 
 where  $\vphi\wedge \psi =\psi-(\vphi-\psi)_-$.
Similar condition holds true for $\Cone^{\iota}$.
\end{description}

By applying Theorem \ref{EquivPI}, we immediately obtain the following corollary.
\begin{coro}\label{MainTh2}
Assume {\bf (A. 1)} -- {\bf (A. 5)}.
The following conditions are equivalent:
\begin{itemize}
\item[{\rm (i)}] $e^{-\beta H} \rhd 0$ w.r.t. $\PI$ for all $\beta >0$.
\item[{\rm (ii)}] $e^{-\beta H^{\iota}(\Lambda)} \rhd 0$ w.r.t. $\PI_{\Lambda}$ for all $\Lambda\in \Bb$ and  $\beta >0$.
\end{itemize}
\end{coro}
Corollary \ref{MainTh2} tells us that  the positivity improvingness of $e^{-\beta H}$ 
is characterized by that of the semigroups generated by the local Hamiltonians,  $H^{\iota}(\Lambda)$.

The following theorem  immediately follows from   Corollary \ref{MainTh2} and  the Perron-Frobenius-Faris theorem \cite{Faris}.

\begin{Thm}\label{UniqG}Assume {\bf (A. 1)} -- {\bf (A. 5)}.
Suppose that $E=\inf \mathrm{spec}(H)$ is an eigenvalue. 
The following conditions are equivalent:
\begin{itemize}

\item[{\rm (i)}]$E$ is a simple eigenvalue. The corresponding eigenvector  can be chosen to be  strictly positive  with respect to $\PI$.
\item[{\rm (ii)}] $e^{-\beta H^{\iota}(\Lambda)} \rhd 0$ w.r.t. $\PI_{\Lambda}$ for all $\Lambda\in \Bb$ and  $\beta >0$.
\end{itemize}
\end{Thm}

\section{Proofs of theorems in Section \ref{DefRe}}\label{Proofs}
\subsection{Properties of the natural cones}
\setcounter{equation}{0}

For each $\Lambda\in \mathbb{B}^d$, we introduce an orthogonal projection $Q_{\Lambda}$ by 
$Q_{\Lambda}=\one^{\iota}_{\Lambda} \otimes P_{\Lambda^c}$, where $P_{\Lambda^c}=|\OU_{\Lambda^c}\ra\la \OU_{\Lambda^c}|$. The operator $Q_{\Lambda}$ will play an important role.
Here, we examine some basic properties of $Q_{\Lambda}$.

Suppose that $\Lambda\subset \Lambda', |\Lambda| \neq 0$ and  $|\Lambda'\setminus  \Lambda|\neq 0$.
We consider a map $\tau_{\Lambda,\Lambda'}:\, \HI_{\Lambda} \to \HI_{\Lambda'}$ defined 
by $\tau_{\Lambda, \Lambda'}(\vphi)=\vphi\otimes \OU_{\Lambda'\setminus  \Lambda}$ for all $\vphi \in  \HI_{\Lambda}$. Trivially, $\tau_{\Lambda, \Lambda'}$ is an isometry. By using this map,  one can regard $\HI_{\Lambda}$ as a  closed subspace of $\HI_{\Lambda'}$, that is,
$\HI_{\Lambda} \cong  \HI_{\Lambda} \otimes \OU_{\Lambda'\setminus  \Lambda}  \subseteq \HI_{\Lambda'}$.
From this point of view, $Q_{\Lambda}$ is the orthogonal projection from $\HI$ onto $\HI_{\Lambda}$.

\begin{Prop}\label{QProp}
One obtains the following:
\begin{itemize}
\item[{\rm (i)}]
 If $\Lambda\subset \Lambda', |\Lambda| \neq 0$ and  $|\Lambda'\setminus  \Lambda|\neq 0$, then $Q_{\Lambda}\le Q_{\Lambda'}$, where the inequality indicates the standard operator inequality\footnote{Let $A$ and $B$ be bounded self-adjoint operators on $\mathfrak{X}$.
Then $A \ge B$ if and only if $\la x|Ax\ra \ge \la x |B x\ra$ for all $x\in \mathfrak{X}$}.
In particular, $Q_{\Lambda}\le \one^{\iota}$.
\item[{\rm (ii)}]  For each $\Lambda\in \mathbb{B}_{\rm b}^d$, $Q_{\Lambda} \unrhd 0$ w.r.t. $\Cone^{\iota}$.
\item[{\rm (iii)}] 
$\displaystyle 
\mbox{{\rm s} -$\displaystyle \lim_{ \Lambda\uparrow \BbbR^d; \ \Lambda\in \mathbb{B}^d_{\rm b}}$
}
Q_{\Lambda}=\one^{\iota}
$, where $\mbox{{\rm s}-$\lim$}$ indicates the strong limit.
\end{itemize}
\end{Prop}
\begin{Proof}
  (i) Because $\MU$ is factorizable, we have 
  $\OU_{\Lambda^c}=\OU_{\Lambda\rq{}\setminus \Lambda}\otimes \OU_{\Lambda'^c}$ by (ii) of Definition \ref{QLS}. Hence, 
   $Q_{\Lambda}$ and $Q_{\Lambda'}$ can be expressed as 
\begin{align}
Q_{\Lambda}=\one^{\iota}_{\Lambda}\otimes P_{\Lambda'\setminus  \Lambda}\otimes P_{\Lambda'^c},\ \ 
Q_{\Lambda'}=\one^{\iota}_{\Lambda}\otimes \one^{\upsilon}_{\Lambda'\setminus  \Lambda}\otimes P_{\Lambda'^c}.
\end{align}
Thus, we have 
\begin{align}
Q_{\Lambda'}-Q_{\Lambda}=\one^{\iota}_{\Lambda}\otimes (\one^{\upsilon}_{\Lambda'\setminus  \Lambda}-P_{\Lambda'\setminus  \Lambda}) \otimes P_{\Lambda'^c} \ge 0.
\end{align}
By taking $\Lambda'=\BbbR^d$, we have $Q_{\Lambda} \le Q_{\BbbR^d}=\one^{\iota}$.

(ii) 
First, recall (iv) of  Definition \ref{QLS2}:  $\MI=\MI_{\Lambda} \otimes \MU_{\Lambda^c}$.
Let $A=\sum_{i=1}^N B_i\otimes C_i$ with $B_i\in \MI_{\Lambda}$ and $C_i\in \MU_{\Lambda^c}$. Trivially, it holds that  $AJ^{\iota}AJ^{\iota} \Omega^{\iota}\ge 0$ w.r.t. $\Cone^{\iota}$, where $\OI$ is the cyclic and separating unit vector for $\MI$ and  $J^{\iota}$ stands for the modular conjugation  associated with $\{\MI, \OI\}$.
Because $A$\rq{}s of this form are dense in $\mathfrak{M}$ under the strong operator topology \cite[Theorem 2.4.11]{BR1}, it suffices to prove that $Q_{\Lambda} AJ^{\iota}AJ^{\iota}\OI\ge 0$ w.r.t. $\PI$.
We have, by  using  (iii) of Definition \ref{QLS2}: $\OI=\OI_{\Lambda} \otimes \OU_{\Lambda^c}$, 
\begin{align}
Q_{\Lambda} AJAJ \OI=\sum_{i, j=1}^N \la \OU_{\Lambda^c}|C_iJ^{\upsilon}_{\Lambda^c}C_j J^{\upsilon}_{\Lambda^c} \OU_{\Lambda^c}\ra
 B_i J^{\iota}_{\Lambda}B_jJ^{\iota}_{\Lambda} \OI_{\Lambda} \otimes \OU_{\Lambda^c}, \label{ExpQ}
\end{align}
where $J_{\Lambda}^{\upsilon}$  is the modular conjugation  associated with  $\{\MU_{\Lambda}, \OU_{\Lambda} \}$.
Let us define a matrix $M=\{M_{ij}\}$ by $
M_{ij}=\la \OU_{\Lambda^c}|C_iJ^{\upsilon}_{\Lambda^c}C_jJ^{\upsilon}_{\Lambda^c}\OU_{\Lambda^c}\ra
$. We claim that $M$ is positive semidifinite.
Indeed, we have
\begin{align}
\sum_{i, j=1}^Nz_iz_j^*M_{ij}=\la \OU_{\Lambda^c}| \bigg[\sum_{i=1}^N z_i C_i \bigg] J^{\upsilon}_{\Lambda^c} \bigg[\sum_{i=1}^N z_i C_i\bigg]J^{\upsilon}_{\Lambda^c} \OU_{\Lambda^c} \ra \ge 0,\ z_1, \dots, z_N\in \BbbC.
\end{align}
Hence, there exists a unitary matrix $U$ such that $M=U \mathrm{diag}(\lambda_1, \dots, \lambda_N) U^*$,
where $\lambda_i$ are     the eigenvalues of $M$. Of course, each $\lambda_i$ is nonnegative.
Inserting this into (\ref{ExpQ}), we have 
\begin{align}
\mbox{the RHS of (\ref{ExpQ})}& =\sum_{k=1}^N \lambda_k
 \bigg[
\sum_{i=1}^N U_{ik} B_i 
\bigg]
J^{\iota}_{\Lambda}
\bigg[
\sum_{i=1}^N U_{ik} B_i 
\bigg]
J^{\iota}_{\Lambda} \OI_{\Lambda} \otimes \OU_{\Lambda^c}\label{ProjA}\\
&=\sum_{k=1}^N \lambda_k D_k J^{\iota}D_kJ^{\iota} \OI, 
\end{align}
where $D_k=\sum_{i=1}^NU_{ik} B_i \otimes \one^{\upsilon}_{\Lambda^c}$. Therefore, we conclude that $Q_{\Lambda}AJ^{\iota}AJ^{\iota} \Omega \ge 0$ w.r.t. $\Cone^{\iota}$.

(iii) We set $\MI_{\mathrm{loc}}=\bigcup_{\Lambda\in \Bb} \MI_{\Lambda}$ and let $
\HI_{\mathrm{loc}} =\MI_{\mathrm{loc}} \OI
$. Then, due to (iv) of Definition \ref{DefMnet},  $\HI_{\mathrm{loc}}$ is dense in $\HI$. For each $\vphi \in \mathfrak{H}_{\mathrm{loc}}$, there exits a $\Lambda\in \Bb$ such that $\vphi\in \HI_{\Lambda}$.
Thus, if we take $\Lambda'$ large as $\Lambda\subset \Lambda'$, then  we have 
$\vphi \cong \vphi\otimes \OU_{\Lambda'\setminus  \Lambda} \otimes \OU_{\Lambda'^c}$, which implies that 
$
Q_{\Lambda'} \vphi\cong \vphi \otimes \OU_{\Lambda'\setminus  \Lambda} \otimes \OU_{\Lambda'^c}=\vphi \otimes \OU_{\Lambda^c}\cong \vphi
$. Accordingly, we obtain $
\mbox{s-$\displaystyle \lim_{\Lambda\uparrow \BbbR^d;\ \Lambda\in \mathbb{B}^d_{\rm b}}$}Q_{\Lambda}\vphi=\vphi 
$. \qed \end{Proof}

\begin{Prop}\label{BasicPro}
We have the following:
\begin{itemize}
\item[{\rm (i)}] If 
$\Lambda\subset \Lambda'$, $\Lambda\in \mathbb{B}_{\rm b}^d$ and  $|\Lambda'\setminus  \Lambda|\neq 0$, then $\Cone^{\iota}_{\Lambda'}=\Cone^{\iota}_{\Lambda}\otimes \Cone^{\upsilon}_{\Lambda'\setminus  \Lambda}$, 
where the tensor product of self-dual cones is defined in Appendix \ref{DefTens}.  In particular, 
 $\Cone^{\iota}=\Cone^{\iota}_{\Lambda}\otimes \Cone^{\upsilon}_{\Lambda^c}$ for all $\Lambda\in \mathbb{B}_{\rm b}^d$.
\item[{\rm (ii)}] If $\Lambda\subset \Lambda', |\Lambda| \neq 0$ and  $|\Lambda'\setminus  \Lambda|\neq 0$, then $\Cone^{\iota}_{\Lambda}\subseteq \Cone^{\iota}_{\Lambda'}$. (More precisely, we have $\Cone^{\iota}_{\Lambda}\otimes \OU_{\Lambda'\setminus  \Lambda} \subseteq \Cone^{\iota}_{\Lambda'}$.)
\item[{\rm (iii)}] If $\Lambda\subset \Lambda', |\Lambda| \neq 0$ and  $|\Lambda'\setminus  \Lambda|\neq 0$, then $Q_{\Lambda}\Cone^{\iota}_{\Lambda'}= \Cone^{\iota}_{\Lambda}
\otimes \OU_{\Lambda'\setminus  \Lambda}$. (More precisely, we have $Q_{\Lambda} \Cone^{\iota}_{\Lambda'} \otimes \OU_{\Lambda'^c} =\Cone^{\iota}_{\Lambda} \otimes \OU_{\Lambda'\setminus  \Lambda} \otimes \OU_{\Lambda'^c}$.)
\item[{\rm (iv)}] $\displaystyle \Cone^{\iota}=\overline{\bigcup_{\Lambda\in \Bb} \Cone^{\iota}_{\Lambda}}$, where the bar denotes the closure  in the norm of $\HI$.
\end{itemize}
\end{Prop}
\begin{Proof} (i) Because $\MI_{\Lambda'}=\MI_{\Lambda} \otimes \MU_{\Lambda'\setminus  \Lambda}$
by (iv) of Definition \ref{QLS2}, we have $\mathcal{P}_0(\MI_{\Lambda'})
=\mathcal{P}_0(\MI_{\Lambda} \otimes \MU_{\Lambda'\setminus  \Lambda})
$. By the definition of the tensor product of self-dual cones in Appendix \ref{DefTens}, we conclude the assertion.

(ii) For each $\vphi\in \Cone^{\iota}_{\Lambda}$, we readily  confirm  that $\vphi\otimes \OU_{\Lambda'\setminus  \Lambda}$ belongs  to $\Cone^{\iota}_{\Lambda'}$. Hence,  $\Cone^{\iota}_{\Lambda} \otimes \OU_{\Lambda'\setminus  \Lambda} \subseteq \Cone^{\iota}_{\Lambda'}$.

(iii) Let $A=\sum_{i=1}^N B_i\otimes C_i$ with $B_i\in \MI_{\Lambda}$ and $C_i\in \MU_{\Lambda'\setminus  \Lambda}$. 
Using  arguments similar to those in the proof of Proposition \ref{QProp} (ii) (or  using \eqref{ProjA}), we obtain $
Q_{\Lambda} AJ^{\iota}_{\Lambda'}A J^{\iota}_{\Lambda'}\OI_{\Lambda'}\in \Cone^{\iota}_{\Lambda}\otimes \OU_{\Lambda'\setminus  \Lambda}$.
This means that $Q_{\Lambda} \mathcal{P}_0(\MI_{\Lambda'}) \OI_{\Lambda'} \subseteq \Cone_{\Lambda} \otimes \OU_{\Lambda'\setminus  \Lambda}$.
Because $\mathcal{P}_0(\MI_{\Lambda'}) \OI_{\Lambda'}$ is dense in $\Cone^{\iota}_{\Lambda'}$, 
we conclude that $Q_{\Lambda}\Cone^{\iota}_{\Lambda'} \subseteq \Cone^{\iota}_{\Lambda} \otimes \OU_{\Lambda'\setminus  \Lambda}$. To prove the converse is easy.

(iv) By applying (ii), we  see that $
\overline{\bigcup_{\Lambda\in \Bb} \Cone^{\iota}_{\Lambda}} \subseteq \Cone^{\iota}
$. We will prove the converse.
Let $\vphi\in \Cone^{\iota}$. Because of (iii) of this proposition,  $Q_{\Lambda} \vphi $ belongs to $\Cone^{\iota}_{\Lambda} \cong \Cone^{\iota}_{\Lambda} \otimes \OU_{\Lambda^c}$ for all $\Lambda\in \Bb$.
By using (iii) of  Proposition \ref{QProp}, we conclude that $\vphi= 
\mbox{s-$\displaystyle \lim_{\Lambda\uparrow \BbbR^d;\ \Lambda\in \mathbb{B}^d_{\rm b}}$}
Q_{\Lambda} \vphi
\in \overline{\bigcup_{\Lambda\in \Bb} \Cone^{\iota}_{\Lambda}}
$. \qed \end{Proof}

\subsection{Some auxiliary lemmas }
In this subsection, we always assume that {\bf (A. 1)}, {\bf (A. 2)} and {\bf (A. 3)}.
Given $\Lambda\in \mathbb{B}^d_{\rm b}$, let $E_{\Lambda}(\cdot)$ be the spectral measure of $W(\Lambda)$.
We set 
\begin{align}
W_ {n}^+(\Lambda)=E_{\Lambda}((-\infty, n])W(\Lambda),\ \ W_{ n}^-(\Lambda)=E_{\Lambda}([-n, \infty))W(\Lambda).
\end{align}
Hence, $W_n^+(\Lambda)$ is bounded from above, and $W_n^-(\Lambda)$ is bounded from below.
\begin{lemm}\label{Lem1}
Let $\Lambda\in \mathbb{B}^d_{\rm b}$.
We have the following.
\begin{itemize}
\item[{\rm (i)}] For each $s\ge 0$, $e^{-s W_{ n}^-(\Lambda)} \in \mathfrak{Z}(\MI)$ and $e^{-s W_{n}^-(\Lambda)} \unrhd 0$ w.r.t. $\PI$.
\item[{\rm (ii)}] For each $s\ge 0$, $e^{s W_{n}^+(\Lambda)} \in \mathfrak{Z}(\MI)$ and $e^{s W_{n}^+(\Lambda)} \unrhd 0$ w.r.t. $\PI$.
\end{itemize}
\end{lemm}
\begin{Proof} (i) 
By using  {\bf (A. 1)} and functional calculus, we know that $e^{-s W_n^-(\Lambda)} \in \mathfrak{Z}(\MI)$.

Let $f(x)=1_{[-n, \infty)}(x) e^{-s x}$, where, given $A\subset \BbbR$,  $1_A$ stands for the indicator function of   $A$. We have $e^{-s W_n^-(\Lambda)} = f(W(\Lambda))$.
By applying Theorem \ref{fPP}, we have $e^{-s W_n^-(\Lambda)} \unrhd 0$ w.r.t. $\PI$ for all $s\ge 0$.

Similarly, we can show (ii). \qed \end{Proof}

\begin{lemm}\label{Lem2} Let $\Lambda\in \mathbb{B}^d_{\rm b}$.
Let $\vphi, \psi\in \PI$. If $\la \vphi|\psi\ra=0$, then we have $\la \vphi|e^{-s W_n^{-}(\Lambda)} \psi\ra=0$ and  $\la \vphi|e^{s W_n^{+}(\Lambda)} \psi\ra=0$ for all $s\ge 0$.
\end{lemm}
\begin{Proof} Applying Lemma \ref{Lem1} and Corollary \ref{fPPC}, we obtain
$
0\le \la \vphi|e^{-s W_n^{-}(\Lambda)} \psi\ra\le e^{s n} \la \vphi|\psi\ra=0.
$
Similarly, we can show $\la \vphi|e^{s W_n^+(\Lambda)}\psi\ra=0$.
\qed \end{Proof}

\begin{lemm}\label{Lem3} Let $\Lambda\in \mathbb{B}^d_{\rm b}$.
One obtains the following.
\begin{itemize}
\item[{\rm (i)}] $L(\Lambda)\dot{+}W_{ n}^-(\Lambda)$ converges to $H$ in the strong resolvent sense as $n\to \infty$.

\item[{\rm (ii)}]  $H\dot{-}W_{ n}^+(\Lambda)$  converges to $L(\Lambda)$ in the strong resolvent sense as $n\to \infty$.
 \end{itemize}
\end{lemm}
\begin{Proof}
(i) Without loss of generality, we may assume that $H^{\iota}(\Lambda)$ and $H$ are positive for all $\Lambda$.
Let $t_n$ be the closed, positive form associated with $L(\Lambda)\dot{+}W_{n}^-(\Lambda)$.
Then the sequence $\{t_n\}_{n=1}^{\infty}$ satisfies $t_1\ge t_2\ge \cdots \ge t_n\ge t_{n+1} \ge  \cdots$.
By using (ii) and (iii) of Definition \ref{ReNetDef},  we see that $Q(t_n)=Q(L(\Lambda))$ for all $n\in \BbbN$. Let $Q(t_{\infty})=\bigcup_{n=1}^{\infty} Q(t_n)=Q(L(\Lambda))$ and 
\begin{align}
t_{\infty}(\vphi, \vphi)=\lim_{n\to \infty} t_n(\vphi, \vphi),\ \ \ \vphi\in Q(t_{\infty}).
\end{align}
Then $H$ is the self-adjoint operator corresponding to $\overline{t}_{\infty}$, the closure of $t_{\infty}$. Applying \cite[Theorem S. 16]{ReSi1}, we conclude (i). 

 Similarly,  we obtain (ii) by applying \cite[Theorem S. 14]{ReSi1}. \qed \end{Proof}

\begin{lemm}\label{PPBasicH}
We have the following.
\begin{itemize}
\item[{\rm (i)}] $e^{-\beta L(\Lambda)} \unrhd 0$ w.r.t. $\PI$ for all $\beta \ge 0$ and $\Lambda\in \mathbb{B}^d_{\rm b}$.
\item[{\rm (ii)}] $e^{-\beta H} \unrhd 0$ w.r.t. $\PI$ for all $\beta \ge 0$.
\end{itemize}
\end{lemm}
\begin{Proof} 
(i)
By {\bf (A. 3)}, we have $e^{-\beta H^{\iota}(\Lambda)} \unrhd 0$ w.r.t. $\PI_{\Lambda}$ and $e^{-\beta H^{\upsilon}(\Lambda^c)} \unrhd 0$ w.r.t. $\PU_{\Lambda^c}$  for all $\beta \ge 0$.
By using the property $\PI=\PI_{\Lambda}\otimes \PU_{\Lambda^c}$ in (i) of  Proposition \ref{BasicPro}  and the definition of  $\PI_{\Lambda}\otimes \PU_{\Lambda^c}$ in Appendix \ref{DefTens},
 we readily confirm that 
   $e^{-\beta L(\Lambda)}=e^{-\beta H^{\iota}(\Lambda)} \otimes e^{-\beta H^{\upsilon}(\Lambda^c)}\unrhd 0$
w.r.t. $\PI$ for all $\beta \ge 0$.

(ii)
By using (i) of  Lemma \ref{Lem3}, we have
\begin{align}
e^{-\beta H}=\slim \mbox{$\mathrm{s}$-$\displaystyle\lim_{m\to\infty}$} \Big(
e^{-\beta L(\Lambda)/m}e^{-\beta W_n^-(\Lambda)/m}
\Big)^m.\label{SemiLim}
\end{align}
By  applying (i),  Lemmas \ref{Lem1} and \ref{ClosedPP}, we conclude that the right hand side of (\ref{SemiLim})
preserves the positivity  w.r.t. $\PI$ for all $\beta \ge 0$. \qed \end{Proof}

\subsection{
 Proof of Theorem \ref{MainTh1}}
(i) $\Longrightarrow$ (ii): We will apply  Faris' idea in \cite{Faris}.
Given $\psi\in \PI\setminus  \{0\}$, we set 
\begin{align}
K(\psi)=\{\vphi\in \PI\, |\, \la \vphi|e^{-\beta L(\Lambda)} \psi\ra=0\, \forall \beta \ge 0\, \forall \Lambda\in \Bb\}.
\end{align}
Our goal is to prove that $K(\psi)=\{0\}$. We remark that the closedness of $K(\psi)$ will be repeatedly
used in the proof.
Let $\vphi\in K(\psi)$. Thus, the vector $\vphi$ satisfies $\la \vphi|e^{-\beta L(\Lambda)} \psi\ra=0$
for all $\beta \ge 0$ and $\Lambda\in \Bb$. 
Note that $e^{-\beta L(\Lambda)}\psi\ge 0$ w.r.t. $\PI$ for all $\beta \ge 0$ by (i) of  Lemma \ref{PPBasicH}.
By Lemma \ref{Lem2}, we have $
\la e^{-sW_n^-(\Lambda)} \vphi|e^{-\beta L(\Lambda)}\psi\ra=0
$ for all $n\in \BbbN,\ s\ge 0,\  \beta \ge 0$ and $\Lambda\in \Bb$, which implies that $
e^{-s W_n^-(\Lambda)} K(\psi) \subseteq K(\psi)
$. Because $e^{-t L(\Lambda)} K(\psi) \subseteq K(\psi)$ for all $t\ge 0$, we have 
$
(e^{-\beta L(\Lambda)/\ell} e^{-\beta W_n^-(\Lambda)/\ell})^{\ell} K(\psi) \subseteq K(\psi)
$ for all $\ell\in \BbbN$. Taking the limit $\ell\to \infty$, we obtain 
$
e^{-\beta (L(\Lambda)\dot{+}W_n^-(\Lambda))} K(\psi)\subseteq K(\psi)
$ for all $n\in \BbbN$ and $\beta \ge 0$ by \cite[Theorem S. 21]{ReSi1}.
Taking the limit $n\to \infty$, we arrive at $e^{-\beta H}K(\psi)\subseteq K(\psi)$ for all $\beta \ge 0$
by (i) of Lemma \ref{Lem3}.
Therefore, for each $\vphi\in K(\psi)$, it holds that $
\la \vphi|e^{-\beta H} \psi\ra=0
$ for all $\beta \ge 0$. By the assumption (i), $\vphi$ must be $0$.
\medskip

(ii) $\Longrightarrow$ (i):
For each $\psi\in \PI\setminus  \{0\}$, we set 
\begin{align}
J(\psi)=\{\vphi\in \PI\, |\, \la \vphi|e^{-\beta H} \psi\ra=0\, \forall \beta \ge 0\}.
\end{align}
Using  arguments similar to those in the previous part, we have 
$e^{-\beta L(\Lambda)} J(\psi)\subseteq J(\psi)$ for all $\beta \ge 0$ and $\Lambda\in \Bb$.
Hence, for every $\vphi\in J(\psi)$, we obtain $\la \vphi|e^{-\beta L(\Lambda)} \psi\ra=0$
for all $\beta \ge 0$ and $\Lambda\in \Bb$. 
By the assumption (ii), $\vphi$
 must be $0$.  
 Hence, for every $\vphi, \psi\in \PI\setminus  \{0\}$, there exists a  $\beta\ge 0$  
 such that $\la \vphi|e^{-\beta H}\psi\ra>0$. \qed

\subsection{Proof of Theorem \ref{MainTh2Ori}}

(i) $\Longrightarrow$ (ii): Let $\Lambda\in \Bb$.
Fix $\psi\in \PI_{\Lambda} \setminus  \{0\}$, arbitrarily and let 
\begin{align}
I(\psi)=\{\vphi\otimes \OU_{\Lambda^c}\, | \, 
\vphi\in \PI_{\Lambda},\ \la \vphi\otimes \OU_{\Lambda^c}|e^{-\beta L(\Lambda)} \psi\otimes \OU_{\Lambda^c}\ra=0\ \forall \beta\ge 0\}.
\end{align}
Let $\vphi\otimes \OU_{\Lambda^c}\in I(\psi)$.
Using  arguments similar to those in the proof of Theorem \ref{MainTh1}, we can prove that 
 $\la \vphi\otimes \OU_{\Lambda^c}
|e^{-\beta H}\psi\otimes \OU_{\Lambda^c}\ra=0$ for all $\beta \ge 0$.
By the assumption (i), $\vphi$ must be $0$. Hence, $I(\psi)=\{0\}$.
Thus, for each $\vphi\in \PI_{\Lambda}\setminus  \{0\}$, there exists a $\beta \ge 0$ such that 
\begin{align}
0<\la \vphi\otimes \OU_{\Lambda^c}|e^{-\beta L(\Lambda)} \psi\otimes \OU_{\Lambda^c}\ra
=\la \vphi|e^{-\beta H^{\iota}(\Lambda)} \psi\ra
\la \OU_{\Lambda^c}|e^{-\beta H^{\upsilon}(\Lambda^c)} \OU_{\Lambda^c}\ra. \label{PIgat}
\end{align}
Because $e^{-\beta H^{\upsilon}(\Lambda^c)}$ is a  positive operator, it holds that $\la \OU_{\Lambda^c}|e^{-\beta H^{\upsilon}(\Lambda^c)} \OU_{\Lambda^c}\ra\ge 0$.
Combining this with \eqref{PIgat}, we get $\la \vphi|e^{-\beta H^{\iota}(\Lambda)} \psi\ra>0$.
To summarize, for each $\vphi, \psi\in \PI_{\Lambda} \setminus  \{0\}$, there exists a $\beta\ge 0$
such that $\la \vphi|e^{-\beta H^{\iota}(\Lambda)} \psi\ra>0$.

\medskip

(ii) $\Longrightarrow$ (i):
Given $\Lambda\in \mathbb{B}^{d}_{\rm b}$, let $q_{\Lambda}=\one^{\iota}_{\Lambda} \otimes |\omega^{\upsilon}_{\Lambda^c}\ra\la \omega^{\upsilon}_{\Lambda^c}|$, where $\omega^{\upsilon}_{\Lambda}$ is given in {\bf (A. 4)}.
By   using arguments similar to those   in the proof of Proposition \ref{QProp}, we can show that $ q_{\Lambda} \unrhd 0$ w.r.t. $\PI$. In addition, we have $q_{\Lambda}^{\perp} \unrhd 0$ w.r.t. $\PI$ by (iii) of {\bf (A. 4)}, where $q_{\Lambda}^{\perp}=\one^{\iota}-q_{\Lambda}$.

Let  $\vphi\in \HI$ be a nonzero vector. We will show that there exits a $\Lambda_o\in \mathbb{B}^d_{\rm b}$ such that, if $\Lambda\in \mathbb{B}^d_{\rm b}$ and $\Lambda_o\subset \Lambda$, then 
$q_{\Lambda}\vphi\neq 0$. By using (iii) of Proposition \ref{QProp}, $Q_{\Lambda} \vphi\to \vphi$ as $\Lambda \uparrow \BbbR^d$.
Hence, for any $\vepsilon>0$, there is a $\Lambda_o\in \mathbb{B}^d_{\rm b}$ such that if $\Lambda\in \mathbb{B}^d_{\rm b}$ and $\Lambda_o \subset \Lambda$, then $\|\vphi-Q_{\Lambda} \vphi\|<\vepsilon$ holds.
By using  the triangle inequality, we find
\begin{align}
\|q_{\Lambda} \vphi\|\ge \|q_{\Lambda} Q_{\Lambda} \vphi\|- \|q_{\Lambda}(\one-Q_{\Lambda})\vphi\|\ge \|q_{\Lambda} Q_{\Lambda}\vphi\|-\vepsilon.\label{ExpsiIn}
\end{align}
By   applying (i) of {\bf (A. 4)} and the fact $\|Q_{\Lambda} \vphi\| \ge \|\vphi\|-\vepsilon$, we get \begin{align}
\|q_{\Lambda} Q_{\Lambda} \vphi\|=\la \omega_{\Lambda^c}|\Omega_{\Lambda^c}\ra \|Q_{\Lambda} \vphi\| \ge \gamma \|Q_{\Lambda} \vphi\| \ge \gamma (\|\vphi\|-\vepsilon),
\end{align}
which implies that, due to \eqref{ExpsiIn},  $\|q_{\Lambda} \vphi\| \ge \gamma\|\vphi\|-(1+\gamma) \vepsilon$ holds.
By choosing $\vepsilon$ such that $
\gamma\|\vphi\|-(1+\gamma) \vepsilon>0
$ and $\Lambda$ sufficiently large, we conclude that $q_{\Lambda} \vphi\neq 0$.

Choose $\vphi, \psi\in \PI\setminus  \{0\}$, arbitrarily.
By using the above claim, there exists  a $\Lambda\in \Bb$ such that $q_{\Lambda}\vphi\neq 0$
and $q_{\Lambda}\psi\neq 0$. 
For such a  $\Lambda$, we have 
\begin{align}
e^{-\beta L(\Lambda)} =e^{-\beta H^{\iota}(\Lambda)} \otimes e^{-\beta H^{\upsilon}(\Lambda^c)}. \label{Tensor}
\end{align}

Because $q_{\Lambda} \unrhd 0$ and $q_{\Lambda}^{\perp} \unrhd 0$ 
w.r.t. $\PI$, we have, by (i) of  Lemma \ref{PPBasicH},
\begin{align}
e^{-\beta L(\Lambda)}  &\ \ \unrhd   q_{\Lambda} e^{-\beta L(\Lambda)}q_{\Lambda}\no
&\underset{(\ref{Tensor})}{=} \la \omega^{\upsilon}_{\Lambda^c}|e^{-\beta H^{\upsilon}(\Lambda^c)} \omega^{\upsilon}_{\Lambda^c}\ra
e^{-\beta H^{\iota}(\Lambda)} \otimes |\omega^{\upsilon}_{\Lambda^c}\ra\la \omega^{\upsilon}_{\Lambda^c}|.
\end{align}
Because $e^{-\beta H^{\upsilon}(\Lambda^c)}$ is a  positive operator and $\ker(e^{-\beta H^{\upsilon}(\Lambda^c)})=\{0\}$,
 it holds that $
\la \omega^{\upsilon}_{\Lambda^c}|e^{-\beta H^{\upsilon}(\Lambda^c)} \omega^{\upsilon}_{\Lambda^c}\ra>0
$  for all $\beta \ge 0$. Hence,  we obtain that 
\begin{align}
\la \vphi|e^{-\beta L(\Lambda)} \psi\ra
\ge \la \omega^{\upsilon}_{\Lambda^c}|e^{-\beta H^{\upsilon}(\Lambda^c)} \omega^{\upsilon}_{\Lambda^c}\ra
\la q_{\Lambda} \vphi|e^{-\beta H^{\iota}(\Lambda)} q_{\Lambda}\psi\ra.\label{InqL}
\end{align}
By the assumption (ii) in Theorem \ref{MainTh2Ori}, there exists a $\beta\ge 0$
such that the RHS of (\ref{InqL}) is strictly positive. By applying  Theorem \ref{MainTh1}, we conclude (i). \qed

\section{The Nelson net}\label{NelsonNet}\label{NelsonI}
\subsection{Definition of the Nelson model}
\setcounter{equation}{0}

In this section, we will provide the  first example of the renormalized Hamiltonian net.
The {\it Nelson Hamiltonian at a  fixed total momentum } $P\in \BbbR^3$ is defined by 
\begin{align}
H_{\kappa}(P)=\frac{1}{2}(P-\Pf)^2-g\int_{\BbbR^3} dk \frac{1_{B_{\kappa}}(k)}{\sqrt{\vepsilon(k)}}
(a(k)+a(k)^*) +\Hf-E_{\kappa}. \label{Hkappa}
\end{align} 
The operator $H_{\kappa}(P)$ acts on $\mathfrak{H}=\Fock(L^2(\BbbR^3))$, where 
$\Fock(\mathfrak{h})$ is the Fock space over $\mathfrak{h}$:
$
\Fock(\mathfrak{h})=\bigoplus_{n=0}^{\infty} \mathfrak{h}^{\otimes_{\mathrm{s}} n}
$. 
$\Pf, \Hf$ and $E_{\kappa}$ are defined by (\ref{MomOp}), (\ref{FieldOp}) and (\ref{ReEn}), respectively.
By the Kato-Rellich theorem \cite[Theorem X.12]{ReSi2}, $H_{\kappa}(P)$ is self-adjoint on $\D(\Pf^2) \cap \D(\Hf)$, bounded from below,   for each $g>0,\ \kappa>0, m\ge 0$ and $P\in \BbbR^3$.\footnote{
First, recall the following well-known bound: 
\begin{align*}
\|(a(f)+a(f)^*)\vphi\| \le 2 \|\vepsilon^{-1/2} f\|\|(\Hf+\one)^{1/2} \vphi\|,\ f\in \D(\vepsilon^{-1/2}),\ \ \vphi\in \D(\Hf^{1/2}).
\end{align*}
Let $h=1_{B_{\kappa}}/\sqrt{\vepsilon}$. Because $\vepsilon^{-1/2} h\in L^2(\BbbR^3)$ for  {\it all} $m\ge 0$ and $\kappa>0$, we find, by using the above bound, 
\begin{align*}
\|(a(h)+a(h)^*)\vphi\| \le 2 \|\vepsilon^{-1/2} h\| \|(\Hf+\one)^{1/2} \vphi\|\le \eta \|\vepsilon^{-1/2} h\|^2+ \|(\Hf+\one)^{1/2} \vphi\|^2/\eta
\end{align*}
for any  $\eta>0$ and $\vphi\in \D(\Hf)$. Here, we have used the elementary inequality $2ab\le \eta a^2+b^2/\eta$ for $a, b\ge 0$ and $\eta>0$. This bound indicates that the interaction term in \eqref{Hkappa} is infinitesimally small with respect to the free Hamiltonian $(P-\Pf)^2/2+\Hf$ for {\it all}
$g>0, \kappa>0, m\ge 0$ and $P$. Hence, we can apply the Kato-Rellich theorem. 
}

Let $H_{\mathrm{ren}}(P)$ be the renormalized Hamiltonian defined in Theorem \ref{NelsonEx}.
Recall that $H_{\kappa}(P)$ converges to $H_{\rm ren}(P)$ in the strong resolvent sense as $\kappa\to \infty$.
In what follows, we will construct a renormalized Hamiltonian net associated with $H_{\mathrm{ren}}(P)$.

\subsection{Properties of the Fock spaces}
Let  $\mathfrak{h}$ be a complex separable Hilbert space.
Given self-adjoint operator $A$ on  $\mathfrak{h}$, its second quantization, $\dG(A)$, 
is defined by
\begin{align}
\dG(A)=0\oplus \Bigg[
\bigoplus_{n=1}^{\infty} \sum_{j=1}^n \one \otimes \cdots \otimes\underbrace{ A }_{j^{{\rm th}}}\otimes \cdots\otimes  \one
\Bigg].
\end{align}
$\dG(A)$ acts on $\Fock(\mathfrak{h})$ and  is essentially self-adjoint. 
In what follows, 
we denote its closure by the same symbol.

We denote by $a(f)$ the annihilation operator on $\Fock(\mathfrak{h})$ with test vector $f\in \mathfrak{h}$ \cite[Section X. 7]{ReSi2}.
By definition, $a(f)$ is densely defined, closed, and antilinear in $f$.
The adjoint, $a(f)^*$,  is called the creation operator.
The creation- and annihilation operators satisfy the following commutation relations:
\begin{align}
[a(f), a(g)^*]=\la f|g\ra,\ \ \ [a(f), a(g)]=0
\end{align}
on  suitable  domains.

If $\mathfrak{h}=L^2(\Lambda)$ with $\Lambda\in \mathbb{B}^3$, then $a(f)$ and $a(f)^*$ are formally expressed as 
\begin{align}
a(f)=\int_{\Lambda} dk f(k)^* a(k),\ \ \ a(f)^*=\int_{\Lambda} dk f(k)a(k)^*,
\end{align}
where the kernel operators $a(k)$ and $a(k)^*$ satisfy (\ref{CCRs}).
In addition,  if $F$ is the  multiplication operator by a real-valued continuous function $F$ on $\mathbb{R}^3$, then $\dG(F)$ is formally expressed as 
\begin{align}
\dG(F)=\int_{\Lambda} dk F(k)a(k)^*a(k).
\end{align}
In this study, we will occasionally use these convenient expressions.

Recall the following factorization properties of the  Fock space:
\begin{align}
\Fock(\mathfrak{h}_1\oplus \mathfrak{h}_2) =\Fock(\mathfrak{h}_1)\otimes \Fock(\mathfrak{h}_2). \label{Tensor1}
\end{align}
Corresponding to (\ref{Tensor1}), we have the following:
\begin{align}
a(f\oplus g)&=a(f)\otimes \one +\one \otimes a(g),\label{Tensor2}\\
\dG(A_1\oplus A_2)&=\overline{\dG(A_1)\otimes  \one +\one \otimes \dG(A_2)}, \label{Tensor3}
\end{align}
where $\overline{X}$ indicates the closure of the operator $X$, and   the identity \eqref{Tensor2} holds on suitable dense subspaces.\footnote{ E.g., the incomplete tensor product of the finite particle subspaces, the domain of number operator $\dG(\one \oplus \one), $  and so on.}
Let $\omega$ be the Fock vacuum in $
\Fock(\mathfrak{h}_1\oplus \mathfrak{h}_2)
$, and let $\omega_i$ be the Fock vacuum in $
\Fock(\mathfrak{h}_i),\ i=1,2
$. Under the identification (\ref{Tensor1}), we have
\begin{align}
\omega=\omega_1\otimes \omega_2. \label{Factome}
\end{align}

For each $\Lambda\in \mathbb{B}^3$ with $|\Lambda| \neq 0$ and $|\Lambda^c|\neq 0$, we have the decomposition: $L^2(\BbbR^3)=L^2(\Lambda)\oplus L^2(\Lambda^c)$. Using this and (\ref{Tensor1}), we have 
\begin{align}
\mathfrak{H}=\mathfrak{H}_{\Lambda} \otimes \mathfrak{H}_{\Lambda^c}, \label{FactH}
\end{align}
where $\mathfrak{H}_{\Lambda}=\Fock(L^2(\Lambda))$. Similarly, we can check that 
\begin{align}
\mathfrak{H}_{\Lambda'}=\mathfrak{H}_{\Lambda}\otimes \mathfrak{H}_{\Lambda' \setminus  \Lambda}, \label{FactH2}
\end{align} provided that $\Lambda\subset \Lambda'$,   $|\Lambda|\neq 0$ and $|\Lambda\rq{}\setminus \Lambda|\neq 0$.
Remark that $\mathfrak{H}_{\Lambda}$ can be expressed as 
\begin{align}
\mathfrak{H}_{\Lambda}=\bigoplus_{n=0}^{\infty} L^2_{\mathrm{sym}}(\Lambda^{\times n}),  \label{FockL2}
\end{align}
where
\begin{align}
L^2_{\mathrm{sym}}(\Lambda^{\times n}) =\big\{
F\in L^2(\Lambda^{\times n})\, |\, F(k_1, \dots, k_n)=
F(k_{\sigma(1)}, \dots, k_{\sigma(n)})\ \mbox{a.e. $\forall \sigma\in \mathfrak{S}_n$}
\big\}.
\end{align}
Here,  $\mathfrak{S}_n$ is the permutation group on the  set $\{1, \dots, n\}$.

By applying (\ref{Tensor2}) and (\ref{Tensor3}), we obtain the following:
\begin{itemize}
\item[1.] For each $f\in L^2(\BbbR^3)$, 
\begin{align}
a(f)=a(f_{\Lambda}) \otimes \one_{\Lambda^c}+\one_{\Lambda} \otimes a(f_{\Lambda^c}), \label{aDec}
\end{align}
holds on suitable dense subspaces, where $f_{\Lambda}= f 1_{\Lambda}$.
\item[2.] 
Given a  function $F(k)$ on $\BbbR^3$, 
let $F$ be the  multiplication operator by   the function $F(k)$.
Suppose that $F$ is real-valued and continuous. Then we have
\begin{align}
\dG(F)=\overline{\dG(F_{\Lambda}) \otimes \one_{\Lambda^c}+\one_{\Lambda} \otimes \dG(F_{\Lambda^c})}, \label{dGDec}
\end{align}
where $F_{\Lambda}=F1_{\Lambda}$.
\end{itemize}

\subsection{Construction of a generalized local system}\label{ConstGQL}

For each $\Lambda\in \mathbb{B}^3$ with $|\Lambda| \neq 0$,  we set 
\begin{align}
L^{\infty}_{\mathrm{sym}}(\Lambda^{\times n}) =\big\{
F\in L^{\infty}(\Lambda^{\times n})\, |\, F(k_1, \dots, k_n)=
F(k_{\sigma(1)}, \dots, k_{\sigma(n)})\ \mbox{a.e. $\forall \sigma\in \mathfrak{S}_n$}
\big\}.
\end{align}
$
L^{\infty}_{\mathrm{sym}}(\Lambda^{\times n})
$ can be regarded as a von Neumann algebra of multiplication operators acting on $L_{\mathrm{sym}}^2(\Lambda^{\times n})$.
Given $\Lambda\in \mathbb{B}^3$, define
\begin{align}
\mathfrak{M}_{\Lambda}=\bigoplus_{n= 0}^{\infty}
L^{\infty}_{\mathrm{sym}}(\Lambda^{\times n})
\end{align}
  with $
  L^{\infty}_{\mathrm{sym}}(\Lambda^{\times 0})=\BbbC
  $.   (For each $\Lambda\in \mathbb{B}^3$ with $|\Lambda|=0$, we set $\mathfrak{M}_{\Lambda}=\varnothing$.) 
  In what follows, we set 
  \begin{align}
  \mathfrak{M}=\mathfrak{M}_{\BbbR^3}.
  \end{align}
 $\mathfrak{M}$ is a von Neumann algebra on $\mathfrak{H}$,
   and $\mathfrak{M}_{\Lambda}$ is a   von Neumann algebra on $\mathfrak{H}_{\Lambda}$. 
   We readily confirm that $\mathfrak{M}$ admits a local structure $\{\mathfrak{M}_{\Lambda}\}_{\Lambda\in \BbbR^3}$.
Using the identification \eqref{Tensor1}, we have
\begin{align}
\mathfrak{M}_{\Lambda\rq{}}=\mathfrak{M}_{\Lambda}\otimes  \mathfrak{M}_{\Lambda\rq{}\setminus \Lambda},\label{AlgTensor}
\end{align}
provided that $\Lambda \subset \Lambda\rq{}$,  $|\Lambda|\neq 0$  and $|\Lambda\rq{}\setminus \Lambda| \neq 0$.

Let $\xi\in L^2(\BbbR^3)$. Suppose that $\xi(k)>0$ a.e. $k$. We define a normalized vector $\Omega$ in 
$\mathfrak{H}$ by 
\begin{align}
\Omega=c\bigoplus_{n=0}^{\infty} \frac{1}{\sqrt{n!}} \xi^{\otimes n},
\end{align}
  where $c$ is the  normalization constant: $c=e^{-\|\xi\|^2/2}$.
Similarly, for each $\Lambda\in \mathbb{B}^3$, we set
  $\Omega_{\Lambda}=c_{\Lambda}\bigoplus_{n=0}^{\infty} \frac{1}{\sqrt{n!}} \xi_{\Lambda}^{\otimes n}$, where  $c_{\Lambda}=e^{-\|\xi_{\Lambda}\|^2/2}$ with $\xi_{\Lambda}=1_{\Lambda} \xi$.
By using the identification (\ref{FactH}), we have 
\begin{align}
\Omega_{\Lambda'}=\Omega_{\Lambda} \otimes \Omega_{\Lambda'\setminus  \Lambda},
\ \ \Omega=\Omega_{\Lambda}\otimes \Omega_{\Lambda^c}, 
\label{VecTensor}
\end{align}
 provided that $\Lambda\subset \Lambda'$ with   $|\Lambda|\neq 0$ and  $|\Lambda'\setminus  \Lambda| \neq 0$. 
 Because $\xi>0$ w.r.t. $L^2(\BbbR^3)_+$, we find that $\Omega>0$ w.r.t. $\Cone$. Similarly, because 
 $\xi_{\Lambda}>0$ w.r.t.  $L^2(\Lambda)_+$, we have $\Omega_{\Lambda}>0$ w.r.t. $\Cone_{\Lambda}$. Hence, $\Omega$ (resp. $\Omega_{\Lambda}$) is cyclic and separating for $\mathfrak{M}$ (resp. $\mathfrak{M}_{\Lambda}$). These indicate that the basic assumptions of Definition \ref{QLS} are actually satisfied.

\begin{Prop}\label{AssConf}
$(\mathfrak{M}, \mathfrak{M})$ is a generalized local system.
\end{Prop}
\begin{Proof}
Taking \eqref{FactH}, \eqref{AlgTensor} and \eqref{VecTensor} into account, we readily confirm that $\mathfrak{M}$
 is factorizable. Hence, by applying  Proposition \ref{Simple}, we conclude the desired  assertion in the proposition.
 \qed  
\end{Proof}

Let $\Delta$ and $J$ be the modular operator and modular conjugation associated with the pair $\{\mathfrak{M}, \Omega\}$. Trivially, $\Delta=\one$ and $J$ is the complex conjugation on $\mathfrak{H}:
 \ J \vphi=\bigoplus_{n=0}^{\infty}\vphi_n^*$.
Similarly, given $\Lambda\in \mathbb{B}^d$, 
let $\Delta_{\Lambda}$ and $J_{\Lambda}$ be the modular operator and modular conjugation associated with $\{\mathfrak{M}_{\Lambda}, \Omega_{\Lambda}\}$. Then
  $\Delta_{\Lambda}=\one_{\Lambda}$ and $J_{\Lambda}$ is the complex conjugation on $\mathfrak{H}_{\Lambda}$. 
In this case, the corresponding natural cones are  respectively given by 
\begin{align}
\Cone=\bigoplus_{n= 0}^{\infty} L_{\mathrm{sym}}^2(\BbbR^{3n})_+,\ \ 
\Cone_{\Lambda}=\bigoplus_{n= 0}^{\infty} L_{\mathrm{sym}}^2(\Lambda^{\times n})_+, \label{DefConeI}
\end{align}
where 
\begin{align}
L^2_{\mathrm{sym}}(\Lambda^{\times n})_+=
\{ F\in L^2_{\mathrm{sym}}(\Lambda^{\times n})\, |\, F(k_1, \dots, k_n) \ge 0\ \ {\rm a.e.} \}
\end{align}
with $
L^2_{\mathrm{sym}}(\Lambda^{\times 0})_+=\BbbR_+=\{r\in \BbbR\, |\, r\ge 0\}
$.
  The self-dual cone  $\Cone$ is referred to as the {\it Fr\"ohlich cone} \cite{Miyao3}, see also \cite{JFroehlich1,JFroehlich2}.

\begin{lemm}\label{A-4}
Given $\Lambda\in \mathbb{B}^3$ with $|\Lambda|\neq 0$, let $\omega_{\Lambda}$ be the Fock vacuum in $\mathfrak{H}_{\Lambda}:\ \omega_{\Lambda}=1\oplus 0\oplus 0 \oplus \cdots$.
We set  $\omega=\omega_{\BbbR^3}$,   the Fock vacuum in $\h$. Then the net $\{\omega_{\Lambda}\}_{\Lambda\in \mathbb{B}^3}$ satisfies {\bf (A. 4)}.
\end{lemm}
\begin{Proof}
Using  \eqref{Factome}, we have $\omega_{\Lambda'}=\omega_{\Lambda} \otimes \omega_{\Lambda'\setminus  \Lambda}$, provided that $\Lambda\subset \Lambda'$ with $|\Lambda|\neq 0$ and $|\Lambda'\setminus  \Lambda| \neq 0$.
Thus, (ii) of {\bf (A. 4)} is satisfied. By the definitions of $\omega_{\Lambda}$ and $\Cone_{\Lambda}$,
we have $\omega_{\Lambda} \ge 0$ w.r.t. $\Cone_{\Lambda}$ for each $\Lambda\in \mathbb{B}^3$.
In addition, we readily confirm that $\la\omega_{\Lambda}|\Omega_{\Lambda}\ra=c_{\Lambda}\ge c> 0$. Hence, (i) of {\bf (A. 4)} is satisfied.

In \cite{Miyao4, Miyao5}, we already confirmed  that (iii) of {\bf (A. 4)} holds.
For  readers\rq{} convenience, we provide the idea of  the proof.
Using the identification (\ref{FockL2}), we know that 
$
\one_{\Lambda} \otimes |\omega_{\Lambda^c}\ra\la \omega_{\Lambda^c}|=\bigoplus_{n=0}^{\infty} 1_{\Lambda}^{\otimes n}
$ holds, where $1_{\Lambda}$ indicates the multiplication operator by the function $1_{\Lambda}$. Thus, we obtain that 
\begin{align}
\one-\one_{\Lambda} \otimes |\omega_{\Lambda^c}\ra\la \omega_{\Lambda^c}|
=\bigoplus_{n=0}^{\infty}[\one_n-1_{\Lambda}^{\otimes n}], \label{Iden}
\end{align}
where $\one_n$ is the identity operator on $(L^2(\BbbR^3))^{\otimes n}$.
Because $
\one_n-1_{\Lambda}^{\otimes n}\ge 0
$ a.e. as a function on $\BbbR^{3n}$,  we can conclude that the RHS of (\ref{Iden}) preserves the positivity. \qed
\end{Proof}

\begin{rem}\upshape
As we proved in Lemma \ref{A-4}, $\omega$ and $\omega_{\Lambda}$ satisfy {\bf (A. 4)}.  However, these
vectors are neither cyclic nor separating for $\mathfrak{M}$ or $\mathfrak{M}_{\Lambda}$.
 On the other hand, $\Omega$ and $\Omega_{\Lambda}$ do not satisfy {\bf (A. 4)}. These facts illustrate why we need two kinds of vectors $\Omega$ and $\omega$ in Section \ref{SecMainResult}.
\end{rem}

\begin{lemm}\label{A-5}
{\bf (A. 5)} is satisfied. 
\end{lemm}
\begin{Proof}
Given  $\vphi, \psi\in \Cone_{\Lambda}$, we have
$
\vphi\wedge \psi=\bigoplus_{n=0}^{\infty} \vphi_n\wedge \psi_n,
$ where 
\begin{align}
(\vphi_n\wedge \psi_n)(k_1, \dots, k_n)=\min\{
\vphi_n(k_1, \dots, k_n),\ \psi_n(k_1, \dots, k_n)
\}.
\end{align}
 Because $\vphi_n, \psi_n\in L_{\rm sym}^2(\Lambda^{\times n})_+$, we see that 
$(\vphi_n\wedge \psi_n)(k_1, \dots, k_n) \ge 0$ a.e., which implies that $
\vphi\wedge \psi \ge 0
$ w.r.t. $\Cone_{\Lambda}$. \qed
\end{Proof}

\subsection{Construction of the Nelson net}

For notational simplicity, we set $H=H_{\mathrm{ren}}(P)$.
For each $\Lambda \in \mathbb{B}^3_{\mathrm{b}}$, we define the Hamiltonian with an ultraviolet cutoff $\Lambda$ by 
\begin{align}
H^{\iota}(\Lambda)=\frac{1}{2}(P-P_{\mathrm{f}, \Lambda})^2-g\int_{\BbbR^3}dk\frac{1_{\Lambda}(k)}{\sqrt{\varepsilon(k)}}
(a(k)+a(k)^*)+H_{\mathrm{f}, \Lambda}-E(\Lambda),
\end{align}
where $
P_{\mathrm{f}, \Lambda},\ H_{\mathrm{f}, \Lambda}
$ and $E(\Lambda)$ are defined by (\ref{DPL}), (\ref{DHL}) and (\ref{DEng}), respectively. 
Because $\Lambda$ is bounded, we can choose $\kappa>0$ so that $\Lambda \subset B_{\kappa}$.
For such a  $\kappa$, we set 
\begin{align}
H_{\kappa}^{\upsilon}(\Lambda^c)=\frac{1}{2}P_{\mathrm{f}, \Lambda^c}^2-g\int_{\BbbR^3}dk\frac{1_{B_{\kappa}\setminus\Lambda}(k)}{\sqrt{\varepsilon(k)}}
(a(k)+a(k)^*)+H_{\mathrm{f}, \Lambda^c}-E(B_{\kappa}\setminus \Lambda). 
\end{align}
$H^{\iota}(\Lambda)$  acts on $\mathfrak{H}_{\Lambda}$ and is self-adjoint
on $\D(
P_{\mathrm{f}, \Lambda}^2
) \cap 
\D(
H_{\mathrm{f}, \Lambda}
)
$,
  bounded from below. Similarly, $H_{\kappa}^{\upsilon}(\Lambda^c)$ acts on $\mathfrak{H}_{\Lambda^c}$ and is self-adjoint
on $\D(
P_{\mathrm{f}, \Lambda^c}^2
) \cap 
\D(
H_{\mathrm{f}, \Lambda^c}
)
$,
  bounded from below.
Using  (\ref{aDec}) and (\ref{dGDec}), we have, as an operator identity, 
\begin{align}
H_{\kappa}(P)=H^{\iota}(\Lambda)+W(\Lambda)+H^{\upsilon}_{\kappa}( \Lambda^c), \label{KappaDec}
\end{align}
where 
\begin{align}
W(\Lambda)=-(P-P_{{\rm f}, \Lambda}) \cdot P_{{\rm f}, \Lambda^c}.
\end{align}

\begin{lemm}\label{ExHami}
Given  $\Lambda\in \mathbb{B}_{\rm b}^3$,
there exists a self-adjoint operator $H^{\upsilon}(\Lambda^c)$,  bounded from below,  such that $H^{\upsilon}_{\kappa}(\Lambda^c)$ converges to $H^{\upsilon}(\Lambda)$ in the strong resolvent sense as $\kappa\to \infty$.

\end{lemm}
\begin{Proof}  We will apply Nelson\rq{}s idea in \cite{Nelson}.
Let $G_{\kappa}$ be the Gross transformation \cite{EPGross} : $G_{\kappa}=e^{S_{\kappa}}$, where
\begin{align}
S_{\kappa}=\overline{a(F)-a(F)^*},\ \ \ F_{\kappa}=g\frac{1_{B_{\kappa}}-1_{B_K}}{\varepsilon^{1/2}(\varepsilon+k^2/2)},\ \ K>0.\label{DefGr}
\end{align}
Here,  $B_K$ stands for  the closed ball with radius $K$ centered at the origin.
For any linear operator $X$, we set $\tilde{X}=G_{\kappa}XG_{\kappa}^{-1}$.
Choose $K$ sufficiently large such that $\Lambda\subset B_K$,
 Following  Nelson\rq{}s arguments (or see \cite[Proof of Proposition 4.7 (i)]{Miyao5})\footnote{The proofs in \cite{Miyao5, Nelson} can be extended to the massless case: $m=0$.}, we can show that there exists a semibounded self-adjoint operator $\tilde{H}^{\upsilon}(\Lambda^c)$ such that $\tilde{H}^{\upsilon}_{\kappa}(\Lambda^c)$
 converges to $\tilde{H}^{\upsilon}(\Lambda^c)$ in the norm resolvent sense as $\kappa\to \infty$.
 Let $G_{\infty}$ be the unitary operator $G_{\kappa}$ with $\kappa=\infty$. Note that $G_{\kappa}$ strongly converges to $G_{\infty}$ as $\kappa\to \infty$.
  By 
 defining $H^{\upsilon}(\Lambda^c)=G^{-1} \tilde{H}^{\upsilon}(\Lambda)G$, we conclude the desired assertion in the lemma.
\qed \end{Proof}
\medskip

 Combining Lemma \ref{ExHami} with (\ref{KappaDec}), we arrive at the following:
 
\begin{Prop}\label{NelNet}
  For each $\Lambda\in \mathbb{B}_{\mathrm{b}}^3$,  we obtain
 $Q(H^{\iota}(\Lambda)) \cap Q(H^{\upsilon}(\Lambda^c)) \subseteq Q(W(\Lambda))  $ and 
 \begin{align}
 H=H^{\iota}(\Lambda) \dot{+} W(\Lambda) \dot{+}H^{\upsilon}(\Lambda^c).
 \end{align}
  Thus,  the net $\{(H^{\iota}(\Lambda), H^{\upsilon}(\Lambda^c), W(\Lambda))\}_{\Lambda\in \mathbb{B}_{\rm b}^3}$ is a renormalized Hamiltonian net associated with $H=H_{\mathrm{ren}}(P)$.
 \end{Prop}
\begin{Proof}
 We provide a sketch. We employ the notations in the proof of Lemma \ref{ExHami}.
Recall that Nelson proves that there exists a self-adjoint operator $\tilde{H}$, bounded from below, such that  $\tilde{H}_{\kappa}(P)$ converges to $\tilde{H}$ 
in the norm resolvent sense as $\kappa\to \infty$, provided that $K$ is large enough, see, e.g., \cite[Proof of Proposition 4.7 (i)]{Miyao5}.
  By using (\ref{FactH}), (\ref{aDec}) and (\ref{dGDec}), we  can see that 
$Q(\tilde{H}^{\iota}(\Lambda)) \cap Q(\tilde{H}^{\upsilon}(\Lambda^c))=Q(\tilde{H})=
[\cap_{j=1}^3 \D(P_{\mathrm{f}, j})] \cap \D(\Hf^{1/2})
$ and $
Q(\tilde{H}^{\iota}(\Lambda)) \cap Q(\tilde{H}^{\upsilon}(\Lambda^c)) \subseteq Q(\tilde{W}(\Lambda))
$.  Given  self-adjoint operator $A$, let $q_A$ be the quadratic form associated with $A$.
Then one can show that $q_{\tilde{H}}(\Phi, \Psi)=q_{\tilde{H}^{\upsilon}(\Lambda^c)}(\Phi, \Psi)+q_{\tilde{W}(\Lambda)}(\Phi, \Psi)+q_{\tilde{H}^{\iota}(\Lambda)}(\Phi, \Psi)$ for each $\Phi, \Psi\in Q(\tilde{H})$. 
Because   $K$ is  sufficiently large, we see that $q_{\tilde{H}}$ and $q_{\tilde{H}^{\upsilon}(\Lambda^c)}$ are closed and   bounded from below. Hence, $\tilde{H}=\tilde{H}^{\iota}(\Lambda) \dot{+} \tilde{W}(\Lambda) \dot{+}\tilde{H}^{\upsilon}(\Lambda^c)$ holds true.
 \qed \end{Proof}

\begin{define}
{\rm 
The net  $\{(H^{\iota}(\Lambda), H^{\upsilon}(\Lambda^c), W(\Lambda))\}_{\Lambda\in \mathbb{B}_{\rm b}^3}$ defined in Proposition \ref{NelNet} is called the {\it Nelson net}
}.
\end{define}

\begin{rem} \upshape
Let $\Lambda, \Lambda\rq{}\in \mathbb{B}^3_{\rm b}$. 
If $\Lambda\subset \Lambda'$ and  $|\Lambda'\setminus  \Lambda| \neq 0$, then
\begin{align}
H^{\iota}(\Lambda')=H^{\iota}(\Lambda) + W(\Lambda'; \Lambda) + H^{\upsilon}(\Lambda'\setminus  \Lambda), \label{AlgRe}
\end{align}
where $W(\Lambda'; \Lambda)=-(P-P_{\mathrm{f}, \Lambda}) \cdot P_{\mathrm{f}, \Lambda'\setminus  \Lambda}$ and
\begin{align}
H^{\upsilon}(\Lambda\rq{}\setminus \Lambda)=\frac{1}{2}P_{\mathrm{f}, \Lambda\rq{}\setminus \Lambda}^2-g\int_{\BbbR^3}dk\frac{1_{\Lambda\rq{} \setminus\Lambda}(k)}{\sqrt{\varepsilon(k)}}
(a(k)+a(k)^*)+H_{\mathrm{f}, \Lambda\rq{}\setminus \Lambda}-E(\Lambda\rq{}\setminus \Lambda),
\end{align}
This relation is very similar to the one appearing in the theory of quantum spin systems, 
see, e.g., \cite[Section 6.2]{BR2}.
\end{rem}

\begin{lemm}\label{A-1toA-3}
The Nelson net satisfies {\bf (A. 1)}, {\bf (A. 2)} and {\bf (A. 3)}.
\end{lemm}
\begin{Proof}
 {\bf (A. 2)} is  trivial because $\Delta=\one$. Because  the restriction of $(W(\Lambda)+i)^{-1}$ to the $n$-particle space $L^2_{\mathrm{sym}}(\BbbR^{3n})$ is a multiplication operator, 
we obtain  that $(W(\Lambda)+i)^{-1}\in \mathfrak{Z}(\mathfrak{M})=\mathfrak{M}$, 
Hence,  {\bf (A. 1)} holds.
 Finally, we will  prove {\bf (A. 3)}. By applying  arguments  similar to those of \cite[Lemma 4.3]{Miyao5}(or applying Proposition \ref{BasicPertPP}), we can prove $e^{-\beta H^{\iota}(\Lambda)} \unrhd 0$ w.r.t. $\Cone_{\Lambda}$ for all $\beta \ge 0$ and 
   $e^{-\beta H_{\kappa}^{\upsilon}(\Lambda^c)} \unrhd 0$ w.r.t. $\Cone_{\Lambda^c}$ for all $\beta \ge 0$ and $\kappa>0$. Because
   $e^{-\beta H^{\upsilon}(\Lambda^c)}$ is defined via the strong limit of  $e^{-\beta H_{\kappa}^{\upsilon}(\Lambda^c)}$, 
   we conclude that  $e^{-\beta H^{\upsilon}(\Lambda^c)} \unrhd 0$ w.r.t. $\Cone_{\Lambda}$ for all $\beta \ge 0$  by using Lemma \ref{ClosedPP}. This completes the proof of {\bf (A. 3)}.
   \qed
\end{Proof}

The main theorem in this section is the following.

\begin{Thm}\label{PIH(P)}
The semigroup   $e^{-\beta H_{\mathrm{ren}}(P)}$ improves the positivity w.r.t. $\Cone$ for all $\beta>0, g>0, m\ge 0$ and $P\in \BbbR^3$.
\end{Thm}
\begin{Proof} 
In Lemmas \ref{A-4}, \ref{A-5} and \ref{A-1toA-3}, we already confirmed that the assumptions
{\bf (A. 1)}--{\bf (A. 5)} are satisfied.  Hence, all assumptions in Corollary \ref{MainTh2} are satisfied.

By using  arguments similar to those of the proof of \cite[Proposition 4.4]{Miyao5}, one can show  that $e^{-\beta H^{\iota}(\Lambda)} \rhd 0$ w.r.t. $\Cone_{\Lambda} $ for all $\beta>0$ and $\Lambda \in \mathbb{B}^3_{\mathrm{b}}$.\footnote{We readily extend the proofs in \cite{Miyao5}  to the massless case.}
Thus, by applying Corollary \ref{MainTh2}, we obtain the desired result. \qed \end{Proof}

\begin{rem}
{\rm 
\begin{itemize}
\item[1.]
Theorem \ref{PIH(P)} was conjectured by Fr\"ohlich \cite{JFroehlich1,JFroehlich2}. The  first proof was given in \cite{Miyao5}.
In this paper, we provide  a proof from a viewpoint of renormalized  Hamiltonian nets.  
In contrast with this, uniqueness of ground states for models for which  energy renormalization is unnecessary  has been   already well known \cite{JFroehlich1,JFroehlich2,GeLowen,MS,Miyao,Miyao3,Miyao4,Moller2,Sloan, Sloan2, Spohn2}.
We  also remark that, recently \cite{Lampart},  Lampart   provided an alternative proof of Theorem \ref{PIH(P)} based on the method established in \cite{LS}.
\item[2.] The existence of ground states for related models are well-established \cite{JFroehlich1,JFroehlich2,GeLowen,MS,Moller2,Spohn}.
In particular, applying the method in \cite{LMS}, we can prove that $H_{\mathrm{ren}}(P)$
has a ground state,  provided that $|P|<1$ and $m>0$.  In this case, the ground state is unique
and chosen to be strictly positive w.r.t. $\Cone$ by Theorems \ref{UniqG} and \ref{PIH(P)}.

 \item[3.]
It had been conjectured  that the renormalized massless Nelson model(i.e., $m=0$) has no ground states.
Recently, this conjecture is solved in \cite{Dam2}. (As for the model with  ultraviolet cutoff, see \cite{Dam}.)
Note that, for the proof of the absence  of ground states in \cite{Dam2}, the positivity improvingness of $e^{-\beta H_{\mathrm{ren}}(P)}$ is a basic input.

\end{itemize}
}
\end{rem}

\section{The Nelson net II}\label{NelsonNetII}
\subsection{Definition of the model}\label{DefNelIIFib}
\setcounter{equation}{0}
In this section, we will examine the  Nelson model with a  confining potential $V$:
\begin{align}
 H_{\mathrm{Nelson},\kappa}=-\frac{1}{2}\Delta_x-V-g\int_{\BbbR^3}dk\frac{1_{B_{\kappa}}(k)}{\sqrt{\vepsilon(k)}}
\big(e^{ik\cdot x}a(k)+e^{-ik\cdot x}a(k)^*\big)+\Hf-E_{\kappa},
\end{align}
where $\Delta_x$ is the Laplacian on $L^2(\BbbR^3, dx)$.
The operator $ H_{\mathrm{Nelson}, \kappa}$ acts on $L^2(\BbbR^3,  dx)\otimes \Fock(L^2(\BbbR^3))$.
Below, we will explain how $ H_{\mathrm{Nelson}, \kappa}$ relates with the model given in Section \ref{NelsonNet}.
For simplicity, we assume that 
\begin{description}
\item[{\bf (V. 1)}] $V\in L^2(\BbbR^3,  dx)$, or $V\in L^{\infty}(\BbbR^3, dx)$.
\end{description}
Due to {\bf (V. 1)} and \cite[Theorem X. 15]{ReSi2}, $-\frac{1}{2}\Delta_x-V$ is self-adjoint on $\D(-\Delta_x)$, bounded from below.
Note that, as discussed  in Section  \ref{PINelsonNetII}, this assumption can be relaxed.
By the Kato-Rellich theorem, $ H_{\mathrm{Nelson}, \kappa}$ is self-adjoint on $\D(-\Delta_x)\cap \D(\Hf)$ and bounded from below  for all $g>0, m\ge 0$ and $\kappa>0$.
To apply the theory established in Section \ref{DefRe}, we set 
\begin{align}
\mathscr{U}=\mathscr{F}\exp\big(ix\cdot \Pt\big),
\end{align}
where $\Pt$ is the total momentum operator given  in Section \ref{Introduction} and $\mathscr{F}$ is the Fourier transformation on $L^2(\BbbR^3)$:
\begin{align}
(\mathscr{F} f)(p)=(2\pi)^{-3/2}\int_{\BbbR^3}e^{-i x\cdot p}f(x)dx,\ f\in L^2(\BbbR^3, dx).
\end{align}
We define 
\begin{align}
{\bs H}_{\kappa}=\mathscr{U} H_{\mathrm{Nelson}, \kappa}\mathscr{U}^{-1}.
\end{align}
Let $x=(x_1, x_2, x_3)$  be   the position operator for the particle: $(x_j f)(x)=x_jf(x),\ f\in L^2(\BbbR^3, dx)$ and let $\nabla_x=(\frac{\partial}{\partial x_1}, \frac{\partial}{\partial x_2}, \frac{\partial}{\partial x_3})$ be the nabla on $L^2(\BbbR^3, dx)$. Then
\begin{align}
\mathscr{F} x\mathscr{F}^{-1}=-i \nabla_p,\ \ \ \mathscr{F}(-i\nabla_x)\mathscr{F}^{-1}=p, \label{PNTr}
\end{align}
where $p$ and $\nabla_p$ are the triplet of multiplication operators and the nabla on the momentum $L^2$-space: $L^2(\BbbR^3, dp)$, respectively.
Applying \eqref{PNTr},  we readily confirm that 
\begin{align}
{\bs H}_{\kappa}=
\frac{1}{2}(p-\Pf)^2-V(-i\nabla_p)-g\int_{\BbbR^3} dk \frac{1_{B_{\kappa}}(k)}{\sqrt{\vepsilon(k)}}
\big(a(k)+a(k)^*\big)+\Hf-E_{\kappa},   \label{HamiFour}
\end{align}
where $V(-i \nabla_p)$ is defined through the functional calculus.
By Theorem \ref{NelsonEx}, we have the following.
\begin{Prop}\label{ExHamiII}
For each $g>0$ and $m\ge 0$, 
there exists a unique self-adjoint operator ${\bs H}$, bounded from below, such that 
${\bs H}_{\kappa}$ converges to ${\bs H}$ in  the  strong resolvent sense as $\kappa\to \infty$.
\end{Prop}

Again  we emphasize that the infinite  energy renormalization, i.e., $E_{\kappa}\approx -\infty$,  is necessary to define the  Hamiltonian ${\bs H}$.

Note that if $V\equiv 0$, then ${\bs H}_{\kappa}$ and $H_{\kappa}(P)$ are    related as 
\begin{align}
{\bs H}_{\kappa}=\int_{\BbbR^3}^{\oplus} H_{\kappa}(P)dP.
\end{align}
Similarly, we   readily confirm that, if $V\equiv 0$, then
\begin{align}
{\bs H}=\int_{\BbbR^3}^{\oplus} H_{\mathrm{ren}}(P)dP.
\end{align}

\subsection{Basic algebras and self-dual cones}
Define
\begin{align}
 \MI_{\Lambda}=L^{\infty}(\BbbR^3, dp)\otimes \mathfrak{M}_{\Lambda},\ \ \Lambda\in \mathbb{B}^3, \label{ModiM}
\end{align}
where $\mathfrak{M}_{\Lambda}$ is given in Section \ref{NelsonNet}. 
In what follows, we set 
\begin{align}
\MI=\MI_{\BbbR^3}.
\end{align}
Trivially, $\MI$
 is a von Neumann algebra on $\HI=L^2(\BbbR^3,  dp)\otimes \mathfrak{H}$, and 
 $\MI_{\Lambda}$ is a von Neumann  algebra on $\HI_{\Lambda}=L^2(\BbbR^3,  dp)\otimes \mathfrak{H}_{\Lambda}$, where $\mathfrak{H}$ and $\mathfrak{H}_{\Lambda}$ are defined in Section \ref{NelsonNet} as well. 
 Using \eqref{FactH} and \eqref{FactH2},  we have
\begin{align}
\HI_{\Lambda\rq{}}=\HI_{\Lambda} \otimes \mathfrak{H}_{\Lambda'\setminus \Lambda},\ \ \ \HI=\HI_{\Lambda} \otimes \mathfrak{H}_{\Lambda^c}, \label{ModiHil}
\end{align}
provided that $\Lambda \subset \Lambda\rq{}$ with  $|\Lambda|\neq 0$ and $|\Lambda\setminus \Lambda\rq{}|\neq 0$.
Similarly, using \eqref{AlgTensor}, we have
\begin{align}
\MI_{\Lambda\rq{}}=\MI_{\Lambda} \otimes \mathfrak{M}_{\Lambda'\setminus \Lambda},\ \ \ 
\MI=\MI_{\Lambda} \otimes \mathfrak{M}_{\Lambda^c},  \label{ModAlg}
\end{align}
provided that $\Lambda \subset \Lambda\rq{}$ with $|\Lambda|\neq 0$  and $|\Lambda\setminus \Lambda\rq{}|\neq 0$.

Let $\vphi$ be a strictly positive function in $L^2(\BbbR^3,  dp)$. Suppose that $\vphi$ is normalized.
Then $\vphi$ is cyclic and separating for $L^{\infty}(\BbbR^3, dp)$.
Now we set 
\begin{align}
\OI=\vphi\otimes \Omega,\ \ \OI_{\Lambda}=\vphi\otimes \Omega_{\Lambda},
\end{align}
where $\Omega$ and $\Omega_{\Lambda}$ are given in   Section \ref{ConstGQL}.
Note that $\OI$ (resp. $\OI_{\Lambda}$) is cyclic and separating for $\MI$ (resp. $\MI_{\Lambda}$). 
By using \eqref{VecTensor},  we see that 
\begin{align}
\OI_{\Lambda\rq{}}=\OI_{\Lambda} \otimes \Omega_{\Lambda'\setminus \Lambda},\ \ \ \OI=\OI_{\Lambda}\otimes \Omega_{\Lambda^c},\label{ModVec}
\end{align}
provided that $\Lambda \subset \Lambda\rq{}$ with $|\Lambda|\neq 0$  and $|\Lambda\setminus \Lambda\rq{}|\neq 0$.

Due to \eqref{FockL2}, the Hilbert spaces $\HI$ and $\HI_{\Lambda}$ can be expressed as 
\begin{align}
\HI= \bigoplus_{n=0}^{\infty} L^2(\BbbR^3, dp)\otimes L^2_{\mathrm{sym}}(\BbbR^{3n}),\ \ \
\HI_{\Lambda}= \bigoplus_{n=0}^{\infty} L^2(\BbbR^3, dp)\otimes L^2_{\mathrm{sym}}(\Lambda^{\times n}).
\label{DirectSumHil}
\end{align}
Note that the  following identification will be often useful:
\begin{align}
 L^2(\BbbR^3, dp)\otimes L^2_{\mathrm{sym}}(\Lambda^{\times n})&=L^2\big(\BbbR^3; L^2_{\mathrm{sym}}(\Lambda^{\times n})\big),\ \ \Lambda\in \mathbb{B}^3, 
\end{align}
where $L^2(\BbbR^3; \mathfrak{X})$ denotes the space of square integrable $\mathfrak{X}$-valued functions, and we used the following convention:
$L^2(\BbbR^3)=L^2(\BbbR^3; L^2_{\mathrm{sym}}(\Lambda^{\times 0}))$. 

\begin{Prop}
$(\MI, \mathfrak{M})$ is a generalized local system.
\end{Prop}

\begin{Proof}
As we already confirmed in Section \ref{ConstGQL}, $\mathfrak{M}$ is factorizable. By \eqref{ModiHil}, \eqref{ModAlg} and \eqref{ModVec}, 
we know that (ii)-(iv) of  Definition \ref{QLS2} are satisfied. \qed
\end{Proof}

The modular operator and modular conjugation associated with the pair $\{\MI, \OI\}$ are respectively given by 
\begin{align}
J^{\iota}\vphi=\bigoplus_{n=0}^{\infty} \vphi^*_n,\ \ \Delta^{\iota}=\one. \label{TrivialDel}
\end{align}
Similarly, the modular operator and modular conjugation associated with $\{\MI_{\Lambda}, \OI_{\Lambda}\}$ are given as $
J_{\Lambda}^{\iota}=
$ the complex conjugation on $\HI_{\Lambda}$ and $\Delta_{\Lambda}^{\iota}=\one$.
The corresponding natural cones are respectively given by 
\begin{align}
\PI=\bigoplus_{n=0}^{\infty} L^2(
\BbbR^3, dp)_+\otimes L^2_{\mathrm{sym}}(\BbbR^{3n})_+
,\ \ \PI_{\Lambda}=\bigoplus_{n=0}^{\infty} L^2(
\BbbR^3,  dp)_+\otimes L^2_{\mathrm{sym}}(\Lambda^{\times n})_+.  \label{ConeDirectSum}
\end{align}

\begin{lemm}\label{A-4New}
Let $\{\omega_{\Lambda}\}_{\Lambda\in \mathbb{B}^3}$ be the net of Fock vacuums given in Lemma \ref{A-4}.
Then $\{\omega_{\Lambda}\}_{\Lambda}$ satisfies {\bf (A. 4)} with the following correspondence:
\begin{itemize}
\item $\PI_{\Lambda}$ and $\PI$ are defined by \eqref{ConeDirectSum}.
\item $\PU_{\Lambda}=\Cone_{\Lambda}$ and $\PU=\Cone$, where $\Cone_{\Lambda}$ and $\Cone$ are defined by \eqref{DefConeI}.
\end{itemize}
\end{lemm}

\begin{Proof}
(i) and (ii) of {\bf (A. 4)} were already confirmed  in the proof of Lemma \ref{A-4}.

By applying \eqref{Iden}, we have
\begin{align}
\one^{\iota}-\one^{\iota}_{\Lambda} \otimes| \omega_{\Lambda^c}\ra\la \omega_{\Lambda^c}|
=\one_{L^2(\BbbR^3, dp)} \otimes \Big[
\one_{\mathfrak{H}}-\one_{\mathfrak{H}_{\Lambda}} \otimes| \omega_{\Lambda^c}\ra\la \omega_{\Lambda^c}|
\Big], \label{I-IPR}
\end{align}
where $\one_{\mathfrak{X}}$ stands for the identity operator on $\mathfrak{X}$.
As we showed in the proof of Lemma \ref{A-4}, 
$
\one_{\mathfrak{H}}-\one_{\mathfrak{H}_{\Lambda}} \otimes| \omega_{\Lambda^c}\ra\la \omega_{\Lambda^c}|
\unrhd 0
$ w.r.t. $\Cone$.  Hence, the right hand side of \eqref{I-IPR} preserves the positivity
w.r.t. $\PI$. \qed
\end{Proof}

\begin{lemm}\label{A-5New}
{\bf (A. 5)} is satisfied under the correspondence given in Lemma \ref{A-4New}.
\end{lemm}

\begin{Proof}
For each $\vphi, \psi\in \PI_{\Lambda}$, we have
$
\vphi\wedge \psi=\bigoplus_{n=0}^{\infty} \vphi_n\wedge \psi_n,
$ where 
\begin{align}
(\vphi_n\wedge \psi_n)(p, k_1, \dots, k_n)=\min\{
\vphi_n(p, k_1, \dots, k_n),\ \psi_n(p, k_1, \dots, k_n)
\}.
\end{align}
 Because $\vphi_n, \psi_n\in L^2(\BbbR^3, dp)_+\otimes L_{\rm sym}^2(\Lambda^{\times n})_+$, we see that 
$(\vphi_n\wedge \psi_n)(p, k_1, \dots, k_n) \ge 0$ a.e., which implies that $
\vphi\wedge \psi \ge 0
$ w.r.t. $\PI_{\Lambda}$. \qed
\end{Proof}

\subsection{The Nelson net II}

Given $\Lambda\in \mathbb{B}_{\mathrm{b}}^3$, we define the  local Hamiltonian by 
\begin{align}
{\bs H}^{\iota}(\Lambda)=\frac{1}{2}\big(p-P_{\mathrm{f}, \Lambda}\big)^2-V(-i\nabla_p)-g
\int_{\BbbR^3} dk \frac{1_{\Lambda}(k)}{\sqrt{\vepsilon(k)}}
\big(a(k)+a(k)^*\big)+H_{\mathrm{f}, \Lambda}-E(\Lambda),
\end{align}
where $P_{\mathrm{f}, \Lambda}, H_{\mathrm{f}, \Lambda}$ and $E(\Lambda)$ are defined by (\ref{DPL}), (\ref{DHL})
and (\ref{DEng}), respectively. 
The Hamiltonian ${\bs H}^{\iota}(\Lambda)$ acts on $\HI_{\Lambda}$.
By the Kato-Rellich theorem again, ${\bs H}^{\iota}(\Lambda)$ is self-adjoint on $\D(p^2) \cap \D(P_{\mathrm{f}, \Lambda}^2)\cap \D(H_{\mathrm{f}, \Lambda})$ and bounded from below for all $g>0$ and $ m\ge0$.

Choose $\kappa>0$ so that $B_{\kappa} \supset \Lambda$. Then we set 
\begin{align}
{\bs H}_{\kappa}^{\upsilon}(\Lambda^c)=\frac{1}{2}P_{\mathrm{f}, \Lambda^c}^2-g
\int_{\BbbR^3} dk \frac{1_{B_{\kappa}\setminus \Lambda}(k)}{\sqrt{\vepsilon(k)}}
\big(a(k)+a(k)^*\big)+H_{\mathrm{f}, \Lambda^c}-E(B_{\kappa} \setminus \Lambda).
\end{align}
The operator ${\bs H}_{\kappa}^{\upsilon}(\Lambda^c)$ acts on $\HU_{\Lambda^c}$.
By the Kato-Rellich theorem, ${\bs H}_{\kappa}^{\upsilon}(\Lambda^c)$ is self-adjoint on $ \cap \D(P_{\mathrm{f}, \Lambda^c}^2)\cap \D(H_{\mathrm{f}, \Lambda^c})$ and bounded from below for all $g>0$,  $ m\ge0$ and $\kappa >0$.

As before, we have, as an operator identity,
\begin{align}
{\bs H}_{\kappa}={\bs H}^{\iota}(\Lambda)+{\bs W}(\Lambda)+{\bs H}_{\kappa}^{\upsilon}(\Lambda^c),
\end{align}
where
\begin{align}
{\bs W}(\Lambda)=-\big(p-P_{\mathrm{f}, \Lambda}\big)\cdot P_{\mathrm{f},   \Lambda^c}.
\end{align}
Using  arguments similar to those in the proof of Proposition \ref{ExHamiII}, we can prove  that there is a self-adjoint
operator ${\bs H}^{\upsilon}(\Lambda^c)$, bounded from below,  such that 
${\bs H}_{\kappa}^{\upsilon}(\Lambda^c)$ converges to ${\bs H}^{\upsilon}(\Lambda^c)$ in the strong resolvent sense as $\kappa\to \infty$.
In addition, we obtain the following:

\begin{Prop}\label{NelNet2}
  For each $\Lambda\in \mathbb{B}_{\mathrm{b}}^3$,  we obtain
 $Q({\bs H}^{\iota}(\Lambda)) \cap Q({\bs H}^{\upsilon}(\Lambda^c)) \subseteq Q({\bs W}(\Lambda))  $ and 
 \begin{align}
 {\bs H}={\bs H}^{\iota}(\Lambda) \dot{+} {\bs W}(\Lambda) \dot{+}{\bs H}^{\upsilon}(\Lambda^c). \label{BHEQ}
 \end{align}
  Hence,  $\{({\bs H}^{\iota}(\Lambda), {\bs H}^{\upsilon}(\Lambda^c), {\bs W}(\Lambda))\}_{\Lambda\in \mathbb{B}_{\rm b}^3}$ is a renormalized Hamiltonian net associated with ${\bs H}$.
 \end{Prop}
\begin{Proof}
The equality \eqref{BHEQ} is  another expression  of  Nelson\rq{}s theorem \cite{Nelson}. Note that the Nelson\rq{}s idea can be extended to the massless case. \qed 
\end{Proof}
\begin{define}
{\rm 
The net 
$\{({\bs H}^{\iota}(\Lambda), {\bs H}^{\upsilon}(\Lambda^c), {\bs W}(\Lambda))\}_{\Lambda\in \mathbb{B}_{\rm b}^3}$ 
 in Proposition \ref{NelNet2} is called the {\it Nelson net II}, as a matter of convenience. 
}
\end{define}

The following lemma is needed in Section \ref{PINelsonNetII}.

\begin{lemm}\label{NelsonIIA}
The Nelson net II satisfies {\bf (A. 1)}, {\bf (A. 2)} and {\bf (A. 3)}.
\end{lemm}
\begin{Proof}
Because $({\bs W}(\Lambda)+i)^{-1}$ is a direct sum of multiplication operators, we readily confirm that {\bf (A. 1)} is satisfied. Because of  \eqref{TrivialDel}, {\bf (A. 2)} is trivial.  In Lemma \ref{PPLocalH}, we show {\bf (A. 3)}. \qed
\end{Proof}

\subsection{Positivity improvingness  of $e^{-\beta {\bs H}}$}\label{PINelsonNetII}

Our first result in this section is the following:
\begin{Thm}\label{PIEquivII}
Assume {\bf (V. 1)}. In addition, assume the following:
\begin{description}
\item[{\bf (V. 2)}] $\hat{V}(p)\ge 0$ a.e. $p$, where $\hat{V}$ is the Fourier transformation of $V$.
\item[{\bf (V. 3)}] There exists an $\vepsilon>0$ such that $B_{\vepsilon}\subseteq \mathrm{supp}  \hat{V}$, where $B_{\vepsilon}$ is the ball with radius $\vepsilon$ centered at the origin and $\mathrm{supp}\, \hat{V}$ stands for  the support of $\hat{V}$.
\end{description}
The following {\rm (i)} and {\rm (ii)} are equivalent to each other:
\begin{itemize}
\item[{\rm (i)}] $e^{-\beta {\bs H}}$ improves the positivity w.r.t. $\PI$ for all $\beta >0$;
\item[{\rm (ii)}] $e^{-\beta {\bs H^{\iota}(\Lambda)}}$ improves the positivity w.r.t. $\PI_{\Lambda}$
 for all $\beta>0$ and $\Lambda\in \mathbb{B}^3_{\mathrm{b}}$.
\end{itemize}
\end{Thm}
\begin{Proof} 
From Lemmas \ref{A-4New}, \ref{A-5New} and \ref{NelsonIIA}, every assumptions in 
Corollary \ref{MainTh2} are satisfied.
Hence, by applying Corollary \ref{MainTh2}, we obtain the desired result in Theorem \ref{PIEquivII}. \qed \end{Proof}
\medskip

In Appendix \ref{ProofNelsonII}, we show the following proposition:

\begin{Prop}\label{OriNelPro}
Assume {\bf (V. 1)}, {\bf (V. 2)} and {\bf (V. 3)}.
For all $\beta >0, g>0, m\ge 0$ and $\Lambda\in \mathbb{B}^3_{\mathrm{b}}$, the semigroup $e^{-\beta {\bs H}^{\iota}(\Lambda)}$ improves the positivity w.r.t. $\PI_{\Lambda}$.
\end{Prop}

Combining Theorem \ref{PIEquivII} and Proposition \ref{OriNelPro}, we arrive at the following:
\begin{coro}\label{PINelson2Mod}
Assume {\bf (V. 1)}, {\bf (V. 2)} and {\bf (V. 3)}.
The semigroup $e^{-\beta {\bs H}}$ improves the positivity w.r.t. $\PI$ for all $\beta >0,\ g>0$ and $m\ge 0$.
\end{coro}

As we mentioned before, the assumptions {\bf (V. 1)}-{\bf (V. 3)} are rather strong.
We can relax these as follows.
\begin{description}
\item[{\bf (V. 1')}] $V\in L^2(\BbbR^3,  dx)+L^{\infty}(\BbbR^3,  dx)$.
\item[{\bf (V. 4)}] There exists an approximating sequence $\{V_n\}$ for $V$ such that 
the following (i)-(iii) hold:
\begin{itemize}
\item[(i)]  $V_n-V\in L^2(\BbbR^3, dx)$ and $\|V-V_n\|_{L^2}\to \infty$ as $n\to \infty$.
\item[(ii)] For all $n\in \BbbN$ and a.e. $p$, the Fourier transformation  $\hat{V}_n(p)$ exists,  and satisfies
$\hat{V}_n\in L^1(\BbbR^3, dp)$ and $0\le \hat{V}_1(p) \le \cdots \le \hat{V}_n(p) \le \hat{V}_{n+1}(p)\le \cdots $.
\item[(iii)] There exist  an $\vepsilon>0$  and $n_*\in \BbbN$ such that $B_{\vepsilon} \subseteq \mathrm{supp} \hat{V}_{n_*}$.

\end{itemize}
\end{description}
Note that the condition {\bf (V. 1')} guarantees  the self-adjointness of $H_{\mathrm{Nelson}, \kappa}$, see, e.g., \cite[Theorem X. 15]{ReSi2}.
Moreover, Proposition \ref{NelNet2} still holds under this assumption.
It often happens that $\hat{V}$ does not exist, or $\hat{V}$ exists but $\hat{V}\notin L^1(\BbbR^3,  dp)$.
Even in these cases, we can apply our theory on the basis of the assumptions {\bf (V. 4)}.
This is the principal reason for introducing the sequence $\{V_n\}$.
The following example  illustrates  the situation described just before.

\begin{example}
{\rm 
Let us consider the Yukawa potential $\displaystyle 
V(x)=
\frac{e^{-\mu |x|}}{|x|}
$ with $\mu\ge 0$.
Then we have $\displaystyle 
\hat{V}(p)=\frac{2^{1/2}}{p^2+\mu^2}$, which  does not belong to $L^1(\BbbR^3, dp)$.
In this case, we set
\begin{align}
V_n(x)=(2\pi)^{-3/2}\int_{\BbbR^3} e^{ip\cdot x}\hat{V}_n(p)dp,
\end{align}
where 
\begin{align}
\hat{V}_n(p)=
\begin{cases}
\hat{V}(p) & \mbox{if $0\le  \hat{V}(p)\le n$ and $|p| \le n$}\\
n & \mbox{if $n < \hat{V}(p)$}\\
0 & \mbox{otherwise}.
\end{cases}
\end{align}
Then the sequence  $\{V_n\}$ satisfies {\bf (V. 4)}.

}
\end{example}

Now we are ready to state our main result in this section.
\begin{Thm}\label{Extension}
Assume {\bf (V. 1')} and {\bf (V. 4)}. The semigroup $e^{-\beta {\bs H}}$ improves the positivity w.r.t. $\PI$ for all $\beta >0,\  g>0$ and $m\ge 0$.
\end{Thm}
We will provide a proof of Theorem \ref{Extension} in Appendix \ref{Ext2V}.

\begin{rem}
{\rm 
Here, we explain a crucial difference between Theorem \ref{Extension} and the result in \cite{MM}.
It is well-known that  the Fock space $\Fock(L^2(\BbbR^3))$ can be identified with $L^2(Q, d\mu)$, where
 $(Q, \mu)$ is some probability space. Under this identification, the field operator $\phi(f)=\overline{a(f)+a(f)^*} $ with $ f\in L^2(\BbbR^3)$, real-valued,  can be regarded as a Gaussian random variable such that $
 \int_Qd\mu \phi(f) \phi(g)=\la f|g\ra_{L^2}
 $. This representation  is called the {\it Schr\"{o}dinger representation} \cite{Simon2}.
 Recall the definition of $H_{\mathrm{Nelson}}$ in Proposition \ref{NelsonEx}.
In \cite{MM}, Matte and M\o ller proved that $e^{-\beta H_{\mathrm{Nelson}}}$ improves the positivity 
w.r.t. $L^2(\BbbR^3\times Q, dxd\mu)_+$ for all $\beta >0$. Their proof is based on the probability theory.
In contrast with \cite{MM}, Theorem \ref{Extension} holds true in the {\it Fock representation}. Moreover, as we already know, our proof is purely operator theoretic. Therefore,   Theorem \ref{Extension} provides information completely different from \cite{MM}.
}
\end{rem}

\begin{rem}
{\rm 
The conditions on $V$, especially {\bf (V. 2)} and  {\bf (V. 4)}, would be  rather restrictive. 
Indeed, the result in \cite{MM} can be obtained under  more general conditions. In contrast with our results  concerning the Hamiltonian at  fixed total momentum, $H_{\mathrm{ren}}(P)$, the method of  \cite{MM} can cover  $P=0$ only. In this way,  these two methods complement each other and have specific advantages.

}
\end{rem}

\appendix
\section{Some useful results}\label{UsefulT}
\setcounter{equation}{0}

Let $\mathfrak{m}$ be a von Neumann algebra on a  complex separable Hilbert space $\mathfrak{h}$.
Assume that $\mathfrak{m}$ has a cyclic and separating vector $\xi$. Thus, 
$
\mathfrak{h}=\overline{\mathfrak{m} \xi}=\overline{\mathfrak{m}^{\prime} \xi}
$.

We use  $\Delta$ and $J$ to denote  the modular operator and the modular conjugation associated with the pair $\{\mathfrak{m}, \xi\}$.
The natural  cone associated with the pair $\{\mathfrak{m}, \xi\}$ is denoted by $\mathfrak{p}$.

\begin{Thm}\label{fPP}
Let $f\in L^{\infty}(\BbbR)\cap L^1(\BbbR)$. Suppose that $f$ is  nonnegative.
Let $A$ be a self-adjoint operator satisfying the following {\rm (i)} and {\rm (ii)}:
\begin{itemize}
\item[{\rm (i)}] $(A+i)^{-1}\in \mathfrak{Z}(\mathfrak{m})$.
\item[{\rm (ii)}] $\Delta^{it } A\subseteq A\Delta^{it }$ for all $t\in \BbbR$.
\end{itemize}
Then we have $f(A)\unrhd 0$ w.r.t. $\mathfrak{p}$.
\end{Thm}
\begin{Proof} 
Remark that $f(A)\in \mathfrak{Z}(\mathfrak{m})$ by (i) and the functional calculus.
In addition, we have $f(A)\Delta^{-1/4} \subseteq \Delta^{-1/4} f(A)$ by (ii).
By  using \cite[Proposition 2.5.26]{BR1}, we have 
\begin{align}
\mathfrak{p}=\overline{\Delta^{-1/4} \mathfrak{m}'_+\xi}, \label{ConeRep}
\end{align}
where $\mathfrak{m}_+'$ is the set of positive elements of $\mathfrak{m}'$.
For each $B'\in \mathfrak{m}_+'$, we see that 
$
f(A)\Delta^{-1/4}B'\xi=\Delta^{-1/4}f(A)^{1/2} B'f(A)^{1/2}\xi
$. Because $f(A)^{1/2} B' f(A)^{1/2}\in \mathfrak{m}_+'$, we conclude that $f(A)
\mathfrak{p}\subseteq \mathfrak{p}$ from (\ref{ConeRep}).  \qed \end{Proof}

Recall that $\mathfrak{h}_{\rm real}$ stands for the $J$-real subspace: $
 \mathfrak{h}_{\rm real}=\{\vphi\in \mathfrak{h}\, |\, J\vphi=\vphi\}
 $. Let $\vphi \in \mathfrak{h}_{\rm real}$. 
By \cite[Theorem 2.5.28]{BR1},  there  are  $\vphi_{-}, \vphi_+ \in \mathfrak{p}$ such that $\vphi=\vphi_+-\vphi_-$ and  $\la \vphi_+|\vphi_-\ra=0$. With this in mind, we set
\begin{align}
|\vphi|=\vphi_++\vphi_-. \label{Abs}
\end{align}

\begin{coro}\label{fPPC}
Under the assumptions in Theorem \ref{fPP}, we have 
\begin{align}
|\la \vphi|f(A) \psi\ra|\le \|f\|_{\infty}  \la |\vphi|\big||\psi|\ra
\end{align}
 for all $\vphi, \psi\in \mathfrak{h}_{\rm real}$, where $|\vphi|$ and $|\psi|$ are defined by \eqref{Abs}.
 In particular, we have
 \begin{align}
0\le \la \vphi|f(A) \psi\ra\le \|f\|_{\infty} \la \vphi|\psi\ra
\end{align}
for all $\vphi, \psi\in \mathfrak{p}$.
\end{coro}
\begin{Proof} Let $c=\|f\|_{\infty}$. Let $g(x)=c-f(x)$. Clearly $g$ is nonnegative.
For each $n\in \BbbN$, we set $g_n=1_{[-n, n]} g$.
Then $g_n\in L^{\infty}(\BbbR) \cap L^1(\BbbR)$ and $g_n$ is nonnegative for all $n\in \BbbN$.
  Hence, 
$g_n(A) \unrhd 0$ w.r.t. $\mathfrak{p}$ due to  Theorem \ref{fPP}. 
Because $g_n(A)$ strongly converges to $g(A)$ as $n\to \infty$, we obtain $g(A) \unrhd 0$ w.r.t. $\mathfrak{p}$ by applying 
Lemma \ref{ClosedPP}.
 Equivalently,  it holds that $f(A) \unlhd c$ w.r.t. $\mathfrak{p}$.

Let $\vphi, \psi\in \mathfrak{h}_{\rm real}$.
Because $0\unlhd f(A) \unlhd c$ w.r.t. $\mathfrak{p}$, we have 
\begin{align}
|\la \vphi|f(A) \psi\ra|\le \la |\vphi|\big|f(A) |\psi|\ra\le c \la|\vphi|\big||\psi|\ra.
\end{align}
This completes the proof. \qed \end{Proof}
\medskip

For $\vphi, \psi\in \mathfrak{h}_{\rm real}$, 
we set 
\begin{align}
\vphi\wedge \psi=\psi-(\vphi-\psi)_-,\ \ \ \vphi\vee \psi=\vphi+(\vphi-\psi)_+.
\end{align}
 The following lemma will be needed.
\begin{lemm}\label{LatticeLem}
Let  $\vphi, \psi\in \mathfrak{h}_{\rm real}$. We have the following:
\begin{itemize}
\item[{\rm (i)}] $\vphi\wedge\psi=\psi\wedge \vphi$.
\item[{\rm (ii)}] $\vphi\wedge \psi\le \vphi$ and $
\vphi\wedge \psi\le \psi
$ w.r.t. $\mathfrak{p}$.
\item[{\rm (iii)}] Suppose $\vphi, \psi\in \mathfrak{p} $ and $ \vphi\wedge \psi\ge 0$ w.r.t. $\mathfrak{p}$. Then $\la \vphi|\psi\ra=0$ if and only if $\vphi \wedge \psi=0$.
\end{itemize}
\end{lemm}
\begin{Proof}
(i) We observe 
\begin{align}
\vphi\wedge \psi-\psi \wedge \vphi=\psi-(\vphi-\psi)_--\vphi+(\psi-\vphi)_-
=\psi-\vphi-(\psi-\vphi)=0.
\end{align}

(ii) immediately follows from (i).

(iii) Note that 
\begin{align}
\vphi\wedge \psi+\vphi\vee \psi=\vphi+\psi,\ \ \ \|\vphi\wedge\psi\|^2+\|\vphi\vee \psi\|^2
=\|\vphi\|^2+\|\psi\|^2.
\end{align}
Hence, we have $\la \vphi|\psi\ra=\la\vphi\wedge \psi|\vphi\vee \psi\ra$.

Suppose that   $\la \vphi|\psi\ra=0 $. Because $
0\le \vphi\wedge \psi\le \vphi\vee \psi
$ w.r.t. $\mathfrak{p}$, we have
$
0=\la \vphi|\psi\ra=\la\vphi\wedge \psi|\vphi\vee \psi\ra \ge \|\vphi\wedge \psi\|^2
$, which implies that $
\vphi\wedge \psi=0
$.
To prove the converse is easy. \qed

\end{Proof}

\begin{Thm}\label{EquivPI}
Let $A$ be a  positive self-adjoint operator acting on $\mathfrak{h}$.
Suppose that $e^{-tA} \unrhd 0$ w.r.t. $\mathfrak{p}$ for all $t\ge 0$.
Suppose that $\vphi\wedge \psi\ge 0$ w.r.t. $\mathfrak{p}$  for each $\vphi, \psi\in \mathfrak{p}$.
Then the following conditions are equivalent:
\begin{itemize}
\item[{\rm (i)}] $e^{-tA} \rhd 0$ w.r.t. $\mathfrak{p}$ for all $t>0$.
\item[{\rm (ii)}] The semigroup $e^{- tA}$ is ergodic w.r.t. $\mathfrak{p}$.
\end{itemize}
\end{Thm}
\begin{Proof}
This theorem is proved in \cite{Miyao}. For readers' convenience, we provide a proof.

(i) $\Longrightarrow$ (ii): Trivial.

(ii) $\Longrightarrow$ (i): Our proof is based on \cite[Theorem XIII.44]{ReSi4}.

{\bf Step 1.} Let $\vphi, \psi\in \mathfrak{p} \setminus  \{0\}$ and let $B_{\vphi, \psi}=\{t>0\, |\, \la \vphi|e^{-tA} \psi\ra>0\}$. By the assumption, $B_{\vphi, \psi}$ is nonempty.
Fix $s\in B_{\vphi, \psi}$, arbitrarily. Then $\la \vphi|e^{-sA} \psi\ra>0
$. Let $\mu=\vphi\wedge e^{-sA } \psi$. Because $e^{-sA} \unrhd 0$ w.r.t. $\mathfrak{p}$
by the assumption, it holds that $\mu\ge 0$ w.r.t. $\mathfrak{p}$.
 Note that because of (iii) of Lemma \ref{LatticeLem},  $\mu\neq 0$ holds.
In addition, by using  (ii) of  Lemma \ref{LatticeLem},
we have $\mu \le e^{-sA} \psi$ and $\mu\le \vphi$ w.r.t. $\mathfrak{p}$.
Hence, for  every $t>0$,
\begin{align}
\la \vphi|e^{-tA} (e^{-sA} \psi)\ra\ge \la \vphi|e^{-tA}\mu\ra\ge \la \mu|e^{-tA} \mu\ra=\|e^{-tA/2} \mu\|^2>0.
\end{align}
Thus,  $s\in B_{\vphi, \psi}$ and $t>0$ imply $s+t\in B_{\vphi, \psi}$.

{\bf Step 2.} As a function of $t$, $\la \vphi|e^{-tA} \psi\ra$ is analytic in a neighborhood of the interval $(0, \infty)$.
Hence, $(0, \infty) \setminus  B_{\vphi, \psi}$ can have only $0$ as a limit point, otherwise $\la \vphi|e^{-tA} \psi\ra$ is identically $0$. In particular, $B_{\vphi, \psi}$ contains arbitrarily small numbers.
Thus, by {\bf Step 1}, we conclude that $B_{\vphi, \psi}=(0, \infty)$. \qed \end{Proof}

Lastly, we prepare for an abstract lemma:
\begin{lemm}\label{ClosedPP}
 Let $\{A_n\}_{n\in \BbbN}$ be a sequence of bounded operators on   $\mathfrak{h}$. Let $A$ be  a bounded operator on   $\mathfrak{h}$. 
Suppose that $A_n$ weakly  converges to $A$ as $n\to \infty$.
If $A_n\unrhd 0$ w.r.t. $\mathfrak{p}$ for all $n\in \BbbN$, then $A\unrhd 0$ w.r.t. $\mathfrak{p}$.
\end{lemm}
\begin{Proof}
See \cite[Proposition 2.8]{Miyao5}.\qed
\end{Proof}

\section{Tensor products of self-dual cones}\label{DefTens}
\setcounter{equation}{0}

Let $\mathfrak{m}_1$ and $\mathfrak{m}_2$ be von Neumann algebras on  complex separable Hilbert spaces $\mathfrak{h}_1$ and $\mathfrak{h}_2$, respectively.
Suppose that  $\xi_1\in \mathfrak{h}_1$ and $\xi_2 \in \mathfrak{h}_2$ 
are cyclic and separating vectors for $\mathfrak{m}_1$ and $\mathfrak{m}_2$, respectively. 
For $j=1, 2$, the modular operator and the modular conjugation associated with the pair $\{\mathfrak{m}_j, \xi_j\}$ are denoted by $\Delta_j$ and $J_j$.

A vector $\xi_1\otimes \xi_2$ is cyclic and separating for $\mathfrak{m}_1 \otimes \mathfrak{m}_2$ as well. 
We denote by  $\Delta$ and $J$  the modular operator and the modular conjugation  associated with the pair $\{\mathfrak{m}_1\otimes \mathfrak{m}_2, \xi_1\otimes \xi_2\}$, respectively.
We readily check that 
 \begin{align}\Delta=\Delta_1\otimes \Delta_2,\ \ 
 J=J_1\otimes J_2.
 \end{align}
 Here, the conjugation $J_1\otimes J_2$ is defined as follows:
 Let $\Phi\in \mathfrak{h}_1\odot \mathfrak{h}_2$, where $\odot$ indicates  the algebraic tensor product.
Hence, $\Phi$ can be expressed as $
 \Phi=\sum_{i, j=1}^Nc_{ij} \vphi_i\otimes \psi_j
 $, where $c_{ij}\in \BbbC,\ \vphi_i\in \mathfrak{h}_1$ and $\psi_j\in \mathfrak{h}_2$.
 Using this expression, we define $J_1\otimes J_2$ by 
 $
 J_1\otimes J_2\Phi=\sum_{i, j=1}^N  c_{ij}^*  J_1\vphi_i\otimes J_2\psi_j
 $. We can show that $J_1\otimes J_2$ is well-defined and  can be extended to a conjugation on $\mathfrak{h}_1\otimes \mathfrak{h}_2$.

 Let $\mathfrak{p}_1$ and $\mathfrak{p}_2$ be natural cones associated with the pairs $\{\mathfrak{m}_1, \xi_1\}$ and $\{\mathfrak{m}_2, \xi_2\}$, respectively.
 We define a tensor product of $\mathfrak{p}_1$ and $\mathfrak{p}_2$ by the natural cone associated with $\{\mathfrak{m}_1\otimes \mathfrak{m}_2, \xi_1\otimes \xi_2\}$:
\begin{align}
\mathfrak{p}_1\otimes \mathfrak{p}_2:=\overline{\mathcal{P}_0(\mathfrak{m}_1\otimes \mathfrak{m}_2) \xi_1\otimes \xi_2},
\end{align}
where $\mathcal{P}_0(\cdots)$ is defined  in Definition \ref{DefPCone}.

\section{Proof of Proposition \ref{OriNelPro}}\label{ProofNelsonII}
\setcounter{equation}{0}

Again we emphasize that Proposition \ref{OriNelPro} holds in the Fock representation, and 
the standard probabilistic approaches in the Schr\"{o}dinger representation are inapplicable.
The method  in this section is novel and  peculiar  to the Fock representation.

\subsection{Key operator inequalities}\label{ExtToUnBd}
For our purpose, we extend the operator inequalities defined in Section \ref{DefRe} to unbounded operators.

In this subsection, $\mathfrak{K}$ denotes $\mathfrak{H}^{\#}$ or $\mathfrak{H}_{\Lambda}^{\#}\ (\#=\iota, \upsilon)$, and $\mathfrak{Q}$
denotes $\Cone^{\#}$ or $\Cone_{\Lambda}^{\#}\ (\#=\iota, \upsilon)$.
Let $\mathfrak{V}$ be  a dense subspace of $\mathfrak{K}$ such that $\mathfrak{V} \cap \mathfrak{Q}\neq \{0\}$.\footnote{In concrete applications in Sections \ref{NelsonNet} and \ref{NelsonNetII},  $\mathfrak{V}$ satisfies a much stronger condition: $\overline{\mathfrak{V} \cap \mathfrak{Q}}=\mathfrak{Q}$.} Set
\begin{align}
\mathscr{L}(\mathfrak{V})=\{\mbox{$A$: linear operator s.t. $\mathfrak{V}\subseteq \D(A)\cap \D(A^*),\  A\mathfrak{V} \subset \mathfrak{V},\ A^*\mathfrak{V} \subset \mathfrak{V}$}\}.
\end{align}
As before, we denote by $\mathfrak{K}_{\rm real }$ the $J$-real subspace: $\mathfrak{K}_{\rm real}=
\{\vphi\in \mathfrak{K}\, |\, J\vphi=\vphi\}$, where $J$ denotes the conjugation associated with $\mathfrak{Q}$.
The following lemma is easy to check:
\begin{lemm}\label{AuxV}
We have the following:
\begin{itemize}
\item[{\rm (i)}] $\mathscr{L} (\mathfrak{V})$ is a $*$-algebra.
\item[{\rm (ii)}] If $A\in \mathscr{L}(\mathfrak{V})$, then $\D(A) \cap \mathfrak{Q} 
\supseteq  \mathfrak{V}\cap \mathfrak{Q}
\neq \{0\}$.
\item[{\rm (iii)}]
If $A\in \mathscr{L}(\mathfrak{V})$, then $\D(A) \cap \mathfrak{K}_{\rm real} \supseteq  \mathfrak{V}\cap\mathfrak{K}_{\rm real} \neq \{0\}$.\footnote{
Because $\mathfrak{K}_{\rm real }=\mathfrak{Q}-\mathfrak{Q}$, (iii) immediately follows from (ii). 
}
\end{itemize}
\end{lemm}

\begin{define}
{\rm 
\begin{itemize}
\item Let $A\in \mathscr{L}(\mathfrak{V})$. If $A(\D(A) \cap \mathfrak{Q}) \subseteq \mathfrak{Q}$, then we write this as $A\unrhd 0$ w.r.t. $\mathfrak{Q}$. Remark that, due to  (ii) of Lemma \ref{AuxV}, this definition is meaningful.
In this case, we say that $A$ {\it preserves the positivity} w.r.t. $\mathfrak{Q}$.
\item Let $A, B\in \mathscr{L}(\mathfrak{V})$. Suppose that $A(\D(A) \cap \mathfrak{K}_{\rm real}) \subseteq\mathfrak{K}_{\rm real}$ and $B(\D(B) \cap \mathfrak{K}_{\rm real}) \subseteq\mathfrak{K}_{\rm real}$.
If $(A-B)\Big(\D(A) \cap \D(B) \cap \mathfrak{Q}\Big) \subseteq \mathfrak{Q}$, then we write this as $A\unrhd B$ w.r.t. $\mathfrak{Q}$.
\end{itemize}
}
\end{define}

The following proposition is a basic tool.

\begin{Prop}[\cite{Miyao5}]\label{BasicPertPP}
Let $A$ be a positive self-adjoint operator and let $B$ be a  symmetric
 operator. Assume the following:
\begin{itemize}
\item[{\rm (i)}] $B$ is $A$-bounded with relative bound $a<1$, i.e.,
                 $\D(A)\subseteq \D(B)$ and $\|Bx\|\le a \|Ax\|+b\|x\|$
                 for all $x\in \D(A)$.
\item[{\rm (ii)}] $0\unlhd e^{-tA}$ w.r.t. $\mathfrak{Q}$ for all $t\ge 0$.
\item[{\rm (iii)}]$0\unlhd -B$ w.r.t. $\mathfrak{Q}$.
\end{itemize} 
Then $0\unlhd e^{-t(A+B)}$ w.r.t. $\mathfrak{Q}$ for all $t\ge 0$.
\end{Prop}
\begin{Proof} See \cite[Theorem A. 18]{Miyao5}. \qed \end{Proof}
\medskip

Theorem \ref{Mono} below will play an  important role in our proof of Proposition \ref{OriNelPro}.
\begin{Thm}[Monotonicity \cite{Miyao6}]\label{Mono}
Let $A, B$ be self-adjoint positive 
 operators on $\mathfrak{K}$. 
Assume that $B=A-C$ with $C\in \mathscr{B}(\mathfrak{K})$.
Suppose 
 that 
\begin{itemize}
\item[{\rm (i)}] $e^{-\beta A}\unrhd 0$ w.r.t. $\mathfrak{Q}$ for
	     all $\beta\ge 0$;
\item[{\rm (ii)}] $C\unrhd 0$ w.r.t. $\mathfrak{Q}$.
\end{itemize} 
Then we have $e^{-\beta B }\unrhd e^{-\beta A}$
 w.r.t. $\mathfrak{Q}$ for all $\beta\ge  0$. 
\end{Thm} 
\begin{Proof} See \cite[Theorem A. 4]{Miyao5}. \qed \end{Proof}

\subsection{Basic properties of the Nelson net II}
Assume that $V$ satisfies {\bf (V. 1)} and {\bf (V. 2)}. In what follows, we will  focus on the case where $V\in L^2(\BbbR^3, dx)$. Note that we readily apply the arguments below to the case where $V\in L^{\infty}(\BbbR^3, dx)$.

For each $n\in \BbbN$, we set 
\begin{align}
\hat{U}_n(k)
=\begin{cases}
\hat{V}(k) & \mbox{if $\hat{V}(k) \le n$ and $|k|\le n$}\\
n & \mbox{if $n < \hat{V}(k)$ and $|k|\le n$}\\
0 & \mbox{if $n < |k|$.}
\end{cases}
\end{align}
Trivially, $\hat{U}_n$ is bounded and 
\begin{align}
\|
\hat{U}_n
-\hat{V}\|_{L^2} \to 0\ \ \ \mbox{as $n\to \infty$.} \label{L^2Conv}
\end{align}
Now, we define a sequence of potentials $\{U_n\}$ by 
\begin{align}
U_n(x)=(2\pi)^{-3/2} \int_{\BbbR^3} dk e^{ik\cdot x} \hat{U}_n(k).
\end{align}
Note that we employ the notation $U_n$ (instead of more natural notation  $V_n$) in order to avoid  confusion with the  approximating sequence     in {\bf (V. 4)}. Remark that $\|U_n\|_{\infty} <\infty$ for all $n\in \BbbN$.
\begin{lemm}\label{VPPLemm}
We have the following:
\begin{itemize}
\item[{\rm (i)}] $U_n(-i\nabla_p) \unrhd 0$ w.r.t. $L^2(\BbbR^3, dp)_+$ for each $n\in \BbbN$.
\item[{\rm (ii)}]  $V(-i\nabla_p) \unrhd 0$ w.r.t. $L^2(\BbbR^3, dp)_+$.
\end{itemize}
\end{lemm}
\begin{Proof}
Because $e^{ik\cdot(-i \nabla_p)} $ is a translation, we have  $e^{ik\cdot(-i \nabla_p)} \unrhd 0$ w.r.t. $L^2(\BbbR^3, dp)_+$.\footnote{Let $\vphi\in L^2(\BbbR^3, dp)_+$. Then
\[
(e^{ik\cdot (-i\nabla_p)} \vphi)(p)=\vphi(p+k) \ge 0\ \ {\rm a.e.}.
\]
Hence, $
e^{ik\cdot (-i\nabla_p)} \unrhd 0
$ w.r.t. $L^2(\BbbR^3, dp)_+$ for all $k\in \BbbR^3$.
} 
Hence, we obtain 
\begin{align}
U_n(-i\nabla_p)=(2\pi)^{-3/2} \int_{\BbbR^3}dk \underbrace{\hat{U}_n(k)}_{\ge 0} \underbrace{e^{ik\cdot(-i \nabla_p)}}_{\unrhd 0} \unrhd 0\ \ \ \mbox{w.r.t. $L^2(\BbbR^3, dp)_+$.}
\end{align}
Similarly, we have $V(-i\nabla_p)
\unrhd 0
$ w.r.t. $L^2(\BbbR^3,  dp)_+$.
\qed \end{Proof}
\medskip

Before we proceed, we prove a basic lemma:
\begin{lemm}\label{AimplyAI}
Let $\Lambda\in \mathbb{B}^3_{\rm b}$.
Let $A\in \mathscr{B}(\HI_{\Lambda})$. Suppose that $A\unrhd 0$ w.r.t. $\PI_{\Lambda}$.
Then we have the following:
\begin{itemize}
\item[{\rm (i)}] $A\otimes \one_{\Lambda^c} \unrhd 0$ w.r.t. $\PI$, where $\one_{\Lambda^c}$ is the identity operator on $\h_{\Lambda^c}$.
\item[{\rm (ii)}] If $\Lambda\subset \Lambda'$ with $|\Lambda\rq{}\setminus \Lambda|\neq 0$, then  $A\otimes \one_{\Lambda' \setminus  \Lambda} \unrhd 0$ w.r.t. $\PI_{\Lambda'}$, where $\one_{\Lambda\rq{}\setminus \Lambda}$ is the identity operator on $\h_{\Lambda\rq{}\setminus \Lambda}$.
\end{itemize}
\end{lemm}
\begin{Proof} (i) First note that, by (\ref{DefConeI}),
\begin{align}
\PI=\PI_{\Lambda}\otimes \Cone_{\Lambda^c}=\bigoplus_{n=0}^{\infty}
\PI_{\Lambda} \otimes L^2_{\mathrm{sym}} \big(
(\Lambda^c)^{\times n}
\big)_+. \label{ConeDIrectSum2}
\end{align}
The self-dual cone $
\PI_{\Lambda} \otimes L^2_{\mathrm{sym}} \big(
(\Lambda^c)^{\times n}
\big)_+
$
 can be expressed as 
\begin{align}
\PI_{\Lambda} \otimes L^2_{\mathrm{sym}} \big(
(\Lambda^c)^{\times n}
\big)_+=\Big\{
\Psi\in \HI_{\Lambda} \otimes L^2_{\mathrm{sym}}\big(
(\Lambda^c)^{\times n}
\big)\, \Big|\, \Psi(k_1, \dots, k_n) \ge 0\ \ \mbox{w.r.t. $\PI_{\Lambda}$ a.e.}
\Big\},
\end{align}
where we regard $\Psi$ as an $\HI_{\Lambda}$-valued function on $(\Lambda^c)^{\times n}$.
Corresponding to the direct sum decomposition (\ref{ConeDIrectSum2}), 
$A\otimes \one_{\Lambda^c}$ can be written as 
$
A\otimes \one_{\Lambda^c}=\bigoplus_{n=0}^{\infty} A\otimes \one_n,
$
where $\one_n$ is the identity on $L_{\mathrm{sym}}^2\big(
(\Lambda^c)^{\times n}
\big)$. Therefore, it suffices to prove that $A\otimes \one_n \unrhd 0$ w.r.t. 
$
\PI_{\Lambda} \otimes L^2_{\mathrm{sym}} \big(
(\Lambda^c)^{\times n}
\big)_+
$ for each $n$. But this is easy to check. Indeed, we have, for each $\Psi\in \PI_{\Lambda} \otimes L^2_{\mathrm{sym}} \big(
(\Lambda^c)^{\times n}
\big)_+$,
\begin{align}
\big(A\otimes \one_n\Psi\big)(k_1, \dots, k_n)=(A\Psi)(k_1, \dots, k_n) \ge 0\ \ \mbox{w.r.t. $\PI_{\Lambda}$}.
\end{align}

Similarly, we can prove (ii).
\qed \end{Proof}
\medskip

\begin{lemm}\label{VPPLemm2}
Let $\Lambda\in \mathbb{B}^3_{\rm b}$.
One obtains the following:
\begin{itemize}
\item[{\rm (i)}] $U_n(-i\nabla_p) \unrhd 0$ w.r.t. $\PI_{\Lambda}$ for  all $n\in \BbbN$ and $V(-i\nabla_p) \unrhd 0$ w.r.t. $\PI_{\Lambda}$, where we regard  $U_n(-i\nabla_p)$ and $V(-i\nabla_p)$ as operators on $\HI_{\Lambda}$.
\item[{\rm (ii)}] $U_n(-i\nabla_p) \unrhd 0$ w.r.t. $\PI$ for  all $n\in \BbbN$  and $V(-i\nabla_p) \unrhd 0$ w.r.t. $\PI$, where we regard  $U_n(-i\nabla_p)$ and $V(-i\nabla_p)$ as operators on $\HI$.
\end{itemize}
\end{lemm}
\begin{Proof} (i) 
Corresponding to 
 (\ref{ConeDirectSum}), $U_n(-i\nabla_p)$ can be expressed as 
\begin{align}
U_n(-i\nabla_p)=\bigoplus_{\ell=0}^{\infty} U_n(-i\nabla_p)\otimes \one_{\ell},
\end{align}
where $\one_{\ell}$ is the identity operator on $
L^2_{\mathrm{sym}}(\Lambda^{\times \ell})
$.
Let $\Psi\in 
L^2(\BbbR^3)_+\otimes L^2_{\mathrm{sym}}(\Lambda^{\times \ell})_+
$. Trivially, $\Psi(p, k_1, \dots, k_{\ell}) \ge 0$ a.e., which implies, by Lemma \ref{VPPLemm}, that 
\begin{align}
\Big(
U_n(-i\nabla_p)\otimes \one_{\ell} \Psi
\Big)(p, k_1, \dots, k_{\ell}) =\big(
U_n(-i\nabla_p)\Psi
\big)(p, k_1, \dots, k_{\ell}) \ge 0\ \ \mbox{a.e..}
\end{align}
Hence, we conclude that $U_n(-i\nabla_p) \unrhd 0$ w.r.t. $\PI_{\Lambda}$.
Similarly, we have  $V(-i\nabla_p) \unrhd 0$ w.r.t. $\PI_{\Lambda}$.

(ii) Combining Lemma \ref{AimplyAI} and (i),  we immediately get the desired assertion. \qed \end{Proof}
\medskip

In \cite[Proposition 4.4]{Miyao6}, we proved the following useful   lemma: 
\begin{lemm}\label{ErgVnAlln}
There exists an  $n_0\in \BbbN$ such that, for all  $n \ge n_0$, $U_n(-i\nabla_p)$ is ergodic w.r.t. $L^2(\BbbR^3,  dp)_+$. That is, for each $\vphi, \psi\in L^2(\BbbR^3, dp)_+\setminus \{0\}$, there exists an $N\in \BbbN_0=\{0\}\cup \BbbN$ such that 
$\la \vphi|U_n(-i\nabla_p)^N\psi\ra>0$.
\end{lemm}
\begin{Proof}
Due to {\bf (V. 3)}, there exists an $\vepsilon>0$ such that $B_{\vepsilon}\subseteq \mathrm{supp} \hat{V}$.
Hence, taking $n_0\in \BbbN$ sufficiently large, we find that $B_{\vepsilon} \subseteq \mathrm{supp} \hat{U}_{n_0}$. Because $\mathrm{supp} \hat{U}_{n} \subseteq \mathrm{supp} \hat{U}_{n+1}$, we see that $B_{\vepsilon} \subseteq \mathrm{supp} \hat{U}_{n}$, provided that $n \ge n_0$.
By applying arguments similar to those in  the proof of \cite[Proposition 4.4]{Miyao6}, we obtain the assertion  in Lemma \ref{ErgVnAlln}.
\qed
\end{Proof}

For each $n\in \BbbN$ and $\Lambda\in \mathbb{B}^3_{\mathrm{b}}$, we set
\begin{align}
{\bs K}(\Lambda) &=\frac{1}{2}\big(p-P_{\mathrm{f}, \Lambda}\big)^2+H_{\mathrm{f}, \Lambda}-E(\Lambda),\\
{\bs I}_n(\Lambda) &=U_n(-i\nabla_p)+g
\int_{\BbbR^3} dk \frac{1_{\Lambda}(k)}{\sqrt{\vepsilon(k)}}
\big(a(k)+a(k)^*\big)
\end{align}
and ${\bs H}^{\iota}_n(\Lambda)={\bs K}(\Lambda)-{\bs I}_n(\Lambda)$.

\begin{lemm}\label{StrgReSemi}
Let  $\Lambda\in \mathbb{B}_{\rm b}^3$. ${\bs H}^{\iota}_n(\Lambda)$ converges to ${\bs H}^{\iota}(\Lambda)$ in the strong resolvent sense as $n\to \infty$.
\end{lemm}

\begin{Proof} 
By (\ref{L^2Conv}), we have ${\bs H}^{\iota}_n(\Lambda)\psi\to {\bs H}^{\iota}(\Lambda) \psi$
for all $\psi\in C_0^{\infty}(\BbbR^3) \cap \D(P_{\mathrm{f}, \Lambda}^2) \cap \D(H_{\mathrm{f}, \Lambda})$ as $n\to \infty$. Thus, by applying \cite[Theorem VIII. 25 (a)]{ReSi1}, we conclude the desired assertion.
 \qed  \end{Proof}
\medskip

To introduce operator inequalities discussed  in Section \ref{ExtToUnBd},  we define the finite boson subspace of $\HI$ by 
\begin{align}
\HI_{\mathrm{fin}} =\Big\{
\Psi=\{\Psi_n\}\in \HI\, \Big|\, \mbox{There exists an $n\in \BbbN$ such that for all $\ell\ge n$, $\Psi_{\ell}=0$}
\Big\},
\end{align}
where the direct sum representation (\ref{DirectSumHil}) is considered.
Similarly, we can define $\HI_{\Lambda, \mathrm{fin}}$ for each $\Lambda\in \mathbb{B}^3$. 
 In what follows, we employ the following notation:
$ \HI_{\Lambda=\BbbR^3,  \mathrm{fin}}=\HI_{\mathrm{fin}}$.
Note that $\HI_{\Lambda, \mathrm{fin}}$ is dense in $\HI_{\Lambda}$.
Trivially, we have
$a(f) \in \mathscr{L}(\HI_{\Lambda, \mathrm{fin}})$, provided that $f\in L^2(\Lambda)$.  Hence, ${\bs I}_n(\Lambda) \in \mathscr{L}(\HI_{\Lambda, \mathrm{fin}})$ for all $\Lambda\in \mathbb{B}_{\rm b}^3$  and $n\in \BbbN$.

\begin{lemm}\label{PPa}
If $f\in L^2(\Lambda)_+$, then $a(f) \unrhd 0$ and $a(f)^*\unrhd 0$ w.r.t. $\PI_{\Lambda}$.
Thus, ${\bs I}_n(\Lambda) \unrhd 0$ w.r.t. $\PI_{\Lambda}$ for all $n\in \BbbN$.
\end{lemm}
\begin{Proof} By   \cite[Proposition 3.8]{Miyao5},
we have  $a(f) \unrhd 0$ and $a(f)^*\unrhd 0$ w.r.t. $\PI_{\Lambda}$.
Combining this and Lemma \ref{VPPLemm2}, we have ${\bs I}_n(\Lambda) \unrhd 0$ w.r.t. $\PI_{\Lambda}$ for all $n\in \BbbN$.
  \qed \end{Proof}

\begin{lemm}\label{PPSemiHI}
For all $\beta\ge 0, n\in \BbbN$ and $\Lambda\in \mathbb{B}^3_{\mathrm{b}}$, it holds that 
$e^{-\beta {\bs K}(\Lambda)}\unrhd 0$ w.r.t. $\PI_{\Lambda}$.
\end{lemm}
\begin{Proof} Corresponding to the direct sum representation (\ref{DirectSumHil}), we have 
\begin{align}
{\bs K}(\Lambda)=\bigoplus_{\ell=0}^{\infty} {\bs K}_{\ell}(\Lambda),
\end{align}
where
\begin{align}
{\bs K}_{\ell}(\Lambda)=\frac{1}{2}\big(
p-k_1-\cdots-k_{\ell}
\big)^2+\vepsilon(k_1)+\cdots +\vepsilon(k_{\ell})-E(\Lambda)
 \end{align} 
 with ${\bs K}_{\ell=0}(\Lambda)=\frac{1}{2}p^2-E(\Lambda).$
 Here, $k_j$ and $\vepsilon(k_j)$  denote the multiplication operators on $L_{\mathrm{sym}}^2(\BbbR^{3\ell})$
 associated with the functions $k_j$ and $\vepsilon(k_j)$, respectively.
Because $e^{-\beta {\bs K}_{\ell}(\Lambda)}$ is a  multiplication operator associated with 
a  positive function, we have $e^{-\beta{\bs K}_{\ell}(\Lambda)} \unrhd 0$ w.r.t. $
L^2(
\BbbR^3, dp)_+\otimes L^2_{\mathrm{sym}}(\Lambda^{\times \ell})_+
$ for all $\beta\ge 0$. Hence, we obtain the desired result. \qed \end{Proof}

\begin{lemm}\label{PPLocalH}
Let $\Lambda\in \mathbb{B}^3_{\rm b}$.  One obtains the following:
\begin{itemize}
\item[\rm (i)] $e^{-\beta {\bs H}^{\iota}_n(\Lambda)}\unrhd 0$ w.r.t. $\PI_{\Lambda}$ for all $\beta\ge 0 $ and $n\in \BbbN$.
\item[\rm (ii)] $e^{-\beta {\bs H}^{\iota}(\Lambda)}\unrhd 0$ w.r.t. $\PI_{\Lambda}$ for all $\beta\ge 0 $.
\end{itemize}
\end{lemm}
\begin{Proof} 
(i)
By applying  Lemmas \ref{PPa}, \ref{PPSemiHI} and  Proposition \ref{BasicPertPP}, we conclude that 
$e^{-\beta {\bs H}^{\iota}_n(\Lambda)} \unrhd 0$ w.r.t. $\PI_{\Lambda}$ for all $\beta \ge 0$.

(ii) 
Taking 
 Lemma \ref{StrgReSemi} and (i) of this lemma into account, 
we can apply  Lemma \ref{ClosedPP} with $A_n=e^{-\beta {\bs H}^{\iota}_n(\Lambda)}$ and $A=e^{-\beta {\bs H}^{\iota}(\Lambda)}$, and  obtain the desired assertion. 
\qed \end{Proof}

\begin{lemm}\label{InErgodic}
Let $n_0$ be  the natural number given in Lemma \ref{ErgVnAlln}. 
For each $n\in \BbbN$ with $n \ge n_0$ and $\Lambda\in \mathbb{B}^3_{\mathrm{b}}$, ${\bs I}_n(\Lambda)$
is ergodic w.r.t. $\PI_{\Lambda}$. That is, 
for any $\vphi, \psi\in
 \big(\PI_{\Lambda} \cap \mathfrak{H}_{\Lambda, {\rm fin}}^{\iota} \big)
 \setminus  \{0\}$, there exists an $N\in \BbbN_0$ such
 that $\la \vphi|{\bs I}_n(\Lambda)^N\psi\ra >0$.
\end{lemm}
\begin{Proof} Set $A=U_n(-i \nabla_p)$ and $B=a(F)+a(F)^*$ with $F=g1_{\Lambda}/\sqrt{\vepsilon}$.
We already know that $A\unrhd 0$ and $B\unrhd 0$ w.r.t. $\PI_{\Lambda}$.

Take $\vphi=\bigoplus_{j=0}^{\infty} \vphi_j, \psi=\bigoplus_{j=0}^{\infty}\psi_j\in \big( \PI_{\Lambda}\cap \mathfrak{H}_{\Lambda, {\rm fin}}^{\iota}\big)
\setminus  \{0\}$, arbitrarily. It suffices to prove that there exists an $N\in \BbbN_0$ such that
$
\la \vphi|(A+B)^N\psi\ra>0
$. Because $\vphi$ and $ \psi$ are nonzero, there exist  $\ell, m\in \BbbN_0$ such that 
$\vphi_{\ell} \neq 0$ and $\psi_{m}\neq 0$. Hence, we have
$\vphi\ge \vphi_{\ell}$ and $\psi\ge \psi_{m}$ w.r.t. $\PI_{\Lambda}$, where we regard $\vphi_{\ell}$ and $\psi_m$ as vectors in $\HI_{\Lambda}$ in the following manner: 
\begin{align}
\vphi_{\ell} = (0, \dots, 0, \underbrace{\vphi_{\ell}}_{\ell^{\rm th}}, 0, \dots),\ \ \ \psi_{m} = (0, \dots, 0, \underbrace{\psi_{m}}_{m^{\rm th}}, 0, \dots). \label{Idnvhipsi}
\end{align}
Because $(A+B)^{k+\ell+m} \unrhd \binom{k+\ell+m}{k} A^kB^{\ell+m}$ w.r.t. $\PI_{\Lambda}$ for  each $k\in \BbbN_0$ and $B\unrhd a(F)$ w.r.t. $\PI_{\Lambda}$, we find that 
\begin{align}
\la \vphi|(A+B)^{k+\ell+m}\psi\ra &\ge \binom{k+\ell+m}{k} \la \vphi_{\ell}|A^kB^{\ell+m}\psi_{m}\ra\no
&= \binom{k+\ell+m}{k} \la B^{\ell}\vphi_{\ell}|A^kB^m\psi_{m}\ra\no
&\ge \binom{k+\ell+m}{k} \la a(F)^{\ell}\vphi_{\ell}|A^ka(F)^m\psi_{m}\ra\no
&=\sqrt{\ell!m!}\binom{k+\ell+m}{k}  \la f|A^kg\ra,
\end{align}
where 
\begin{align}
f(p)&=\int_{\Lambda^{\times \ell}} dk_1\cdots dk_{\ell} \vphi_{\ell}(p; k_1, \dots, k_{\ell}) F(k_1)\cdots F(k_{\ell}),\\
g(p)&=\int_{\Lambda^{\times m}} dk_1\cdots dk_{m} \psi_{m}(p; k_1, \dots, k_{m}) F(k_1)\cdots F(k_{m}).
\end{align}
Trivially, $f$ and $g$ are nonzero and $f(p)\ge 0$ and $g(p)\ge 0$ a.e. $p$.
Thus, by applying Lemma \ref{ErgVnAlln}, we can choose a  $k\in \BbbN_0$ such that $\la f|A^kg\ra>0$. This completes the proof of Lemma \ref{InErgodic} \qed \end{Proof}
\medskip

\begin{Prop}\label{LlocPI} Let $n_0$ be  the natural number given in Lemma \ref{ErgVnAlln}. 
For all $\beta>0, n\in \BbbN_0$ with $n\ge n_0$ and $\Lambda\in \mathbb{B}^3_{\mathrm{b}}$, we have
$e^{-\beta {\bs H}^{\iota}_n(\Lambda)} \rhd 0$ w.r.t. $\PI_{\Lambda}$.
\end{Prop}
\begin{Proof} The proof of Proposition \ref{LlocPI} is a modification of that of \cite[Proposition 4.4]{Miyao5}.
For readers' convenience, we will provide a  proof here.
Choose $\vphi, \psi\in \PI_{\Lambda}\setminus  \{0\}$, arbitrarily. By (\ref{ConeDirectSum}), we can express $\vphi$ and $\psi$
 as $\vphi=\bigoplus_{n=0}^{\infty} \vphi_n$ and $\psi=\bigoplus_{n=0}^{\infty} \psi_n$ with $\vphi_n, \psi_n\in L^2(\BbbR^3, dp)_+\otimes L^2_{\rm sym}(\Lambda^{\times n})_+$.
 Because $\vphi$ and $\psi$ are non-zero, there exist $n_1, n_2\in \BbbN_0$ such that 
$\vphi_{n_1}\neq 0$ and $\psi_{n_2} \neq 0$.
Since $e^{-\beta {\bs K}(\Lambda)}$ is an injection and preserves the positivity, we have
$e^{-\beta {\bs K}(\Lambda)}\psi_{n_2} \neq 0$ and  $e^{-\beta {\bs K}(\Lambda)}\psi_{n_2} \ge 0$ w.r.t. $\PI_{\Lambda}$.  Here, 
we have used identifications similar to \eqref{Idnvhipsi}.
Hence, by Lemma  \ref{InErgodic}, there exists an $\ell\in \BbbN_0$ such that 
\begin{align}
\big\la\vphi_{n_1}|{\bs I}_n(\Lambda)^{\ell} e^{-\beta {\bs K}(\Lambda)} \psi_{n_2}\big\ra>0.\label{StrictPpsi}
\end{align}

Since $\vphi\ge \vphi_{n_1}$ and $\psi\ge \psi_{n_2}$ w.r.t. $\PI_{\Lambda}$, we get
\begin{align}
\big\la\vphi\big|e^{-\beta {\bs H}^{\iota}_n(\Lambda)} \psi\big\ra  \ge \big\la\vphi_{n_1}\big|e^{-\beta {\bs H}^{\iota}_n(\Lambda)} \psi_{n_2}\big\ra \label{InqCut}
\end{align}
for all $\beta \ge 0$ by Lemma \ref{PPLocalH}. By applying the Duhamel formula\footnote{
For notational simplicity, we set $A={\bs K}(\Lambda)$ and $B={\bs I}_m(\Lambda)$.
On the finite boson subspace $\h_{\Lambda, \mathrm{fin}}^{\iota}$, we have the following under  the strong operator topology:
\[
\frac{d}{ds}e^{-sA}e^{-(\beta-s)(A-B)}=e^{-sA} (-B) e^{-(\beta-s)(A-B)},\ \ 0< s < \beta.
\]
By carrying out the integration, we get 
\[
e^{-\beta(A-B)}=e^{-\beta A}+\int_0^{\beta} e^{-sA}Be^{-(\beta-s)(A-B)}ds
\]
on $\h_{\Lambda, \mathrm{fin}}^{\iota}$. By applying this identity repeatedly, we obtain \eqref{DUHA}.
}
, we have
\begin{align}
e^{-\beta {\bs H}^{\iota}_n(\Lambda)}=\sum_{j=0}^{\ell}D_j+R_{\ell}\ \ \ \mbox{on $\h_{\Lambda, \mathrm{fin}}^{\iota}$}, \label{DUHA}
\end{align}
where $D_0=e^{-\beta {\bs K}(\Lambda)}$ and 
\begin{align}
D_j&=\int_{0\le s_1\le \cdots\le s_j\le \beta} {\bs I}_n(\Lambda)(s_1)\cdots {\bs I}_n(\Lambda)(s_j) e^{-\beta {\bs K}(\Lambda)}ds_1\cdots ds_j, \\
R_{\ell}&= \int_{0\le s_1\le \cdots\le s_{\ell+1}\le \beta} {\bs I}_n(\Lambda)(s_1)\cdots {\bs I}_n(\Lambda)(s_{\ell+1}) e^{-\beta {\bs H}^{\iota}_n(\Lambda)}ds_1\cdots ds_{\ell+1}
\end{align}
with ${\bs I}_n(\Lambda)(s)=e^{-s{\bs K}(\Lambda)} {\bs I}_n(\Lambda)e^{s{\bs K}(\Lambda)}$.
Because ${\bs I}_n(\Lambda) \unrhd 0,\  e^{-\beta {\bs K}(\Lambda)} \unrhd 0$ and $e^{-\beta {\bs H}^{\iota}_n(\Lambda)} \unrhd 0$ w.r.t. $\PI_{\Lambda}$, we obtain 
$D_j\unrhd 0$ and $R_{\ell} \unrhd 0$ w.r.t. $\PI_{\Lambda}$ by Lemma \ref{ClosedPP}.
Hence,
\begin{align}
\big\la\vphi_{n_1}\big|e^{-\beta {\bs H}^{\iota}_n(\Lambda)} \psi_{n_2}\big\ra \ge \big\la\vphi_{n_1}\big| D_{\ell} \psi_{n_2}\big\ra. \label{ExpDell}
\end{align}
Set $G(s_1, \dots, s_{\ell})
=\big\la\vphi_{n_1}\big|
 {\bs I}_n(\Lambda)(s_1)\cdots {\bs I}_n(\Lambda)(s_{\ell}) e^{-\beta {\bs K}(\Lambda)}
\psi_{n_2}\big\ra 
$. By using  (\ref{StrictPpsi}), we have $G(0, \dots, 0)>0$. Because  $G(s_1, \dots, s_{\ell})$ is positive and 
continuous in $s_1, \dots, s_{\ell}$, we find  that 
\begin{align}
\big\la\vphi_{n_1}\big| D_{\ell} \psi_{n_2}\big\ra=\int_{0\le s_1\le \cdots\le s_{\ell}\le \beta} G(s_1, \dots, s_{\ell}) ds_1\cdots ds_{\ell}>0. \label{InqGs}
\end{align}
Combining (\ref{InqCut}), (\ref{ExpDell}) and (\ref{InqGs}), we arrive at 
$
\big\la\vphi\big|e^{-\beta {\bs H}^{\iota}_n(\Lambda)} \psi\big\ra >0
$ for all $\beta>0$. \qed \end{Proof}

\subsection{
 Proof of Proposition \ref{OriNelPro}}
 Let $n_0$ be  the natural number given in Lemma \ref{ErgVnAlln}. 
For each $m, n\in \BbbN$ with $n>m\ge n_0$, we have
\begin{align}
U_{n}(-i\nabla_p)-U_m(-i\nabla_p) =(2\pi)^{-3/2} \int_{\BbbR^3}dk 
\underbrace{\big(
\hat{U}_{n}(k)-\hat{U}_m(k)
\big)}_{\ge 0} 
\underbrace{e^{ik\cdot (-i\nabla_p)}}_{\unrhd 0} \unrhd 0, \label{MonoV}
\end{align}
which implies that ${\bs H}^{\iota}_m(\Lambda) \unrhd {\bs H}^{\iota}_{n}(\Lambda)$ w.r.t. $\PI_{\Lambda}$.
By applying Theorem \ref{Mono}, we obtain
$
e^{-\beta {\bs H}^{\iota}_{n}(\Lambda)} \unrhd e^{-\beta {\bs H}^{\iota}_m(\Lambda)}
$
w.r.t. $\PI_{\Lambda}$ for all $\beta \ge 0$. Taking the limit $n\to \infty$, 
we arrive at 
$
e^{-\beta {\bs H}^{\iota}(\Lambda)} \unrhd e^{-\beta {\bs H}^{\iota}_m(\Lambda)}
$
w.r.t. $\PI_{\Lambda}$ by Lemma \ref{ClosedPP}.
By using  Proposition \ref{LlocPI}, we conclude that $
e^{-\beta {\bs H}^{\iota}(\Lambda)} \unrhd e^{-\beta {\bs H}^{\iota}_m(\Lambda)} \rhd  0
$ w.r.t. $\PI_{\Lambda}$ for all $\beta>0$. \qed

\section{Proof of Theorem \ref{Extension}}\label{Ext2V}
\setcounter{equation}{0}
Let $\{V_n\}$ be the approximating sequence of $V$ given  in {\bf (V. 4)}.
Let ${\bs H}^{(n)}$ be the Hamiltonian ${\bs H}$ with $V$ replaced by $V_n$. 
Using   arguments similar to those in the proof of Lemma \ref{StrgReSemi}, we obtain the following lemma.
\begin{lemm}\label{ResConv}
${\bs H}^{(n)}$ converges to ${\bs H}$ in the norm resolvent sense as $n\to \infty$.
\end{lemm}
\begin{Proof}
 Let $G_{\infty}$ be the Gross transformation defined by (\ref{DefGr}) with $\kappa=\infty$. Set $\tilde{ {\bs H}}=G_{\infty}{\bs H}G_{\infty}^{-1}$ and 
 $\tilde{ {\bs H}}^{(n)}=G_{\infty}{\bs H}^{(n)} G_{\infty}^{-1}$. 
 In addition, set $\mathcal{J}=p^2-V(-i\nabla_p)+\Pf^2+\Hf-c$,
 where $c$ is a constant such that $\mathcal{J}$ is positive.
 Then we readily confirm that $Q(\tilde{{\bs H}}^{(n)})=Q(\tilde{{\bs H}})=Q(\mathcal{J})$ for all $n\in \BbbN$.
 As a form, we have, by using (i) of {\bf (V. 4)}, 
 \begin{align}
 \Big|
 \la \vphi|\tilde{{\bs H}} \vphi\ra-\la \vphi|\tilde{{\bs H}}^{(n)}\vphi\ra
 \Big| &\le \big|
 \la \vphi|\big(
 V(-i\nabla_p)-V_n(-i\nabla_p)
 \big)\vphi\ra
 \big|\no
 &\le \|V-V_n\|_{L^2} \|\vphi\|_{L^2}^2  \to  0\ (n\to \infty)
 \end{align}
  for all $\vphi\in Q(\mathcal{J})$. By applying \cite[Theorem VIII. 25 (c)]{ReSi1},
  we conclude that $\tilde{{\bs H}}^{(n)}$ converges to $\tilde{{\bs H}}$ in the norm resolvent sense.\qed
\end{Proof}

Due to (ii) of {\bf (V. 4)}, $V_n\in L^{\infty}(\BbbR^3, dx)$ holds.
Therefore, we can apply Corollary \ref{PINelson2Mod} and conclude that  
\begin{align}
e^{-\beta {\bs H}^{(n)}} \rhd 0 \ \ \ \mbox{w.r.t. $\PI$ for all $\beta>0$ and $n\ge n_*$}, \label{BasicPIE}
\end{align}
where $n_*$ is the  natural number given in (iii) of {\bf (V. 4)}.

Let ${\bs H}_{\kappa}^{(n)}$ be the Hamiltonian (\ref{HamiFour}) with $V$ replaced by $V_n$. By using   arguments similar to those in \cite{Nelson}, we can show that ${\bs H}^{(n)}_{\kappa}$ converges to ${\bs H}^{(n)}$ in the strong resolvent sense as $\kappa\to \infty$.
For $m, n\in \BbbN$ with $n>m\ge n_*$,  by using (ii) of {\bf (V. 4)} and   arguments similar to those in  the proof of \eqref{MonoV}, we have $V_n(-i\nabla_p)-V_m(-i\nabla_p) \unrhd 0$ w.r.t. $\PI$, which implies that 
\begin{align}
{\bs H}_{\kappa}^{(m)}-{\bs H}_{\kappa}^{(n)}&=V_n(-i\nabla_p)-V_m(-i\nabla_p)\unrhd 0 \ \ \mbox{w.r.t. $\PI$}.
\end{align}
By using Theorem \ref{Mono},  we obtain that 
$e^{-\beta {\bs H}_{\kappa}^{(n)}} \unrhd e^{-\beta {\bs H}_{\kappa}^{(m)}}$ w.r.t. $\PI$ for all $\beta\ge 0$. 
Because ${\bs H}_{\kappa}^{(n)}$ converges to ${\bs H}^{(n)}$ in the strong resolvent sense as $\kappa\to \infty$ by Proposition \ref{ExHamiII},  one obtains $e^{-\beta {\bs H}^{(n)}} \unrhd e^{-\beta {\bs H}^{(m)}}$ w.r.t. $\PI$ for all $\beta\ge 0$ by Lemma \ref{ClosedPP}.
Then taking the limit $n\to \infty$, we get   $e^{-\beta {\bs H}} \unrhd e^{-\beta {\bs H}^{(m)}}$ w.r.t. $\PI$ for all $\beta\ge 0$ by Lemmas \ref{ClosedPP} and  \ref{ResConv}.
Combining this with (\ref{BasicPIE}), we finally obtain  that 
$e^{-\beta {\bs H}} \rhd 0$ w.r.t. $\PI$ for all $\beta> 0$.  \qed

\end{document}